\documentclass[a4paper,12pt]{article}
\usepackage{jheppub}
\usepackage{amsmath,amssymb,bm}\usepackage[T1]{fontenc} 
\usepackage{epsfig}
\usepackage{graphicx}
\usepackage{multirow}
\usepackage{slashed}
\makeatletter

\def\be{\begin{equation}}
\def\ee{\end{equation}}
\def\beq{\begin{equation}}
\def\eeq{\end{equation}}
\def\beqar{\begin{eqnarray}}
\def\eeqar{\end{eqnarray}}
\def\barr{\begin{array}}
\def\earr{\end{array}}

\def\and{\qquad {\rm and } \qquad}

\def\slp{p \hspace{-1ex}/}
\def\sleps{ \epsilon \hspace{-1ex}/}
\def\slk{k \hspace{-1ex}/}

\def\sbar{ \overline{s} }
\def\thmin{\theta_0}

\def\eegz{$e^+e^- \to  Z \gamma~$}

\newcommand\comment[1]{}

\newcommand{\Rmnum}[1]{\expandafter\@slowromancap\romannumeral #1@}
\makeatother

\title{\boldmath New Physics in \eegz
at the ILC with polarized beams:
Explorations beyond conventional anomalous triple gauge boson couplings}

\author[a]{B. Ananthanarayan}
\author[a]{Jayita Lahiri}
\author[b]{Monalisa Patra}
\author[c]{Saurabh D. Rindani}

\affiliation[a]{Centre for High Energy Physics, Indian Institute of Science, Bangalore 560 012, India}
\affiliation[b]{Department of Theoretical Physics, Tata Institute of Fundamental Research, Mumbai}
\affiliation[c]{Theoretical Physics Division, Physical Research Laboratory, Navrangpura, Ahmedabad 380 009, India }

\abstract{
One of the most-studied signals for physics beyond the
standard model in the production of gauge bosons in electron-positron
collisions is that due to the anomalous triple gauge boson couplings in
the $Z \gamma$ final state.  In this work, we study the implications of
this at the ILC with polarized beams for signals that go beyond traditional
anomalous triple neutral gauge boson couplings.  Here we report a dimension-8 
CP-conserving 
$Z\gamma Z$ vertex that has not found mention in the literature.
We carry out a systematic study
of the anomalous couplings in general terms and arrive at a classification.
We then obtain  linear-order distributions with and without CP
violation. Furthermore, we place the study in the context of general BSM
interactions represented by $e^+e^- Z \gamma$ contact interactions. We set up a correspondence
between the triple gauge boson couplings and the four-point contact interactions.
We also present sensitivities on these anomalous couplings, which will
be achievable at the ILC with realistic polarization and luminosity.}

\begin{document}
 \maketitle
 \flushbottom 

\section{Introduction}
The Standard Model (SM) is a well-established theory now and is being tested at very high precision in a 
variety of sectors, e.g., in the Higgs sector at the Large Hadron Collider (LHC), and
in the flavour sector at low-energy 
and high-intensity experiments, to name a couple of examples.
Furthermore, the gauge sector of the SM is predictive and highly constrained. 
The study of gauge-boson pair production will be an important process
to look for new physics at the International Linear Collider (ILC)~\cite{Behnke:2013xla, Baer:2013cma}.  
The ILC is a proposed next generation collider after the LHC 
that will collide electrons and positrons at high energy and luminosity. 
The availability of beam polarization, either longitudinal or
transverse,  of one or both of the beams, 
will also significantly enhance the sensitivity 
to new physics interactions~\cite{MoortgatPick:2005cw, Riemann:2011zzb}.
The rate for gauge-boson pair production
will be sensitive to the gauge-boson self-interactions, which arise through
the non-Abelian nature of the electroweak sector $SU(2)_L \times U(1)_Y$.
Thus, it would be important to look for deviations from SM predictions in this sector.
Nevertheless, gauge invariance and Lorentz invariance as well as renormalizability 
place powerful constraints on the possible structures that can arise.
Thus a model independent classification of terms has been a rich and highly developed 
field, see refs.~\cite{Hag, Renard:1981es, Gounaris:1983zn, Baur:1992cd, Gounaris:1996rz, Gounaris:1999kf}.
The work of Hagiwara et al.~\cite{Hag} will be used by us as a standard touchstone
in the considerations associated with anomalous couplings in the  neutral-boson sector. 

Of the many diboson processes that have been considered, 
$e^+e^-\to Z \gamma$ has received substantial
attention in the past. 
The $Z\gamma Z$ and $Z\gamma \gamma$ couplings
are absent at tree level, and also highly suppressed when allowed 
by internal particle loops in the SM, forbidding
the $s$-channel production of $ZZ$ and $Z\gamma$. 
Therefore any deviation from 
the tree-level SM predictions will signal the presence of beyond-SM
(BSM) physics.
We will first return to the anomalous couplings for this process that 
were introduced some decades ago~\cite{Hag, Baur:1992cd, Gounaris:1999kf}.
In particular, these authors have provided a standard basis, in terms
of 8 couplings, denoted by $h_i^V$, $V=Z,\gamma$, $i=1,2,3,4$, with
$i=1,2$ denoting dimension-6 and -8 CP-violating couplings while
$i=3,4$ denote dimension-6 and -8 CP-conserving couplings.
The individual values of these triple gauge boson couplings
(TGCs) as described before are zero at tree level in the SM, 
with non-zero values
arising at higher orders or in composite models. These anomalous
couplings have been extensively studied in the literature in the context
of different colliders\footnote{While the issue of anomalous triple gauge bosons has been discussed
for several decades now, there have been inequivalent definitions 
in the literature.  For instance, in Ref.~\cite{Czyz:1988yt} it is 
mentioned that
they have a parametrization which is similar to, but not exactly the same as that of 
Hagiwara et al.~\cite{Hag}. The form factors of the two are related
by an overall normalization, with the form factors of Ref.~\cite{Hag} being
($-$2) times those of Ref.~\cite{Czyz:1988yt}. In Ref.~\cite{SDR}
the effective CP-violating Lagrangian has been written down,
and the anomalous couplings are denoted by $\lambda_1$ and $\lambda_2$.
In our work~\cite{Ananthanarayan:2011fr}, we have demonstrated that
these are equivalent to  $f_1$ and $f_2$ of Ref.~\cite{Czyz:1988yt}.}~\cite{Czyz:1988yt, 
SDR, Choi:1994nv, Ellison:1998uy, Baur:2000ae, Atag:2003wm, Ananthanarayan:2011fr, Cata:2013sva, Degrande:2013kka}.
Moreover there has been a lot of work in the literature
\cite{Baur:1992cd, Gounaris:1999kf, Choi:1994nv, Atag:2003wm, Cata:2013sva}
where effective Lagrangians or effective momentum-space vertices and the associated form factors in 
neutral gauge boson production have been discussed.

In all previous work on the subject, there have been no deviations from the set initially
considered by Ref.~\cite{Hag}, in which 
the terms are implicitly symmetric under the 
interchange of $Z \leftrightarrow \gamma$. The lowest-dimension
effective operators within the effective Lagrangian approach for the
neutral anomalous couplings, with all the particles being off-shell,
has been discussed in Ref.~\cite{Larios:1999km, Larios:2000ni}. In that work, 
there is the possibility that there can be terms that 
do not respect this symmetry at the Lagrangian level.  However, we have 
checked that even those terms produce the same anomalous TGCs.
In the present work, we have tried to push this hypothesis further, and indeed 
at dimension-8 we uncover a new term.
Here, we report our finding that
an additional coupling involving only the $Z$ exists, with $Z \gamma Z$ coupling consistent 
with Lorentz invariance, electromagnetic gauge invariance and Bose symmetry, which has not been
explicitly reported in the literature. We consider this to be an important
addition to the body of literature on anomalous TGCs.

Searches for these neutral anomalous couplings have been performed at 
LEP~\cite{Achard:2004ds, Abbiendi:2003va}, 
the Tevatron~\cite{Abazov:2011qp, Aaltonen:2011zc} and
the LHC. The most stringent bounds have come from the 
ATLAS~\cite{Aad:2013izg} and CMS~\cite{Chatrchyan:2013nda}
collaborations, 
with the data taken at $\sqrt{s}$ = 7 TeV. 
Since the anomalous gauge couplings would give rise to photons with
large transverse momentum, $p_T^\gamma$, the LHC collaborations
have placed limits on the couplings by measuring the total production 
cross section and looking at the $p_T$ distribution of the photon.
As the photon transverse energy spectrum has similar sensitivity 
to CP-conserving and CP-violating couplings, the experimental results
are generally in terms of the CP-conserving couplings $h_3^V$ and $h_4^V$.
These analyses are all based on what are claimed to be the most general 
Lorentz invariant effective interactions given by Ref.~\cite{Hag}
\footnote{We have scaled the couplings by the factor
$1/2$ in case of $h_i^Z$ and $1/(4 s_W c_W)$ in case of $h_i^\gamma$
of Ref.~\cite{Hag} for reasons to be explained in the next section.}.
The limits are as follows :
\begin{itemize}
\item ATLAS : $|h_3^Z| < 0.028,~~|h_4^Z| < 1.74 \times 10^{-4},
 ~|h_3^\gamma| < 0.027,~~|h_4^\gamma| < 1.58 \times 10^{-4}$~\cite{Aad:2013izg}
\item CMS : $|h_3^Z| < 5.4 \times 10^{-3},~~|h_4^Z| < 2.6 \times 10^{-5},
~|h_3^\gamma| < 4.9 \times 10^{-3},~~|h_4^\gamma| < 2.5 \times
10^{-5}$~\cite{Chatrchyan:2013nda}
\end{itemize}

It has been pointed out by the authors of~\cite{BASDR1, BASDR2, Abraham:1993zh, Abraham:1998cm}
that one economical way of fingerprinting BSM
physics is to use model-independent contact interactions.
In the present work, we approach the question of studying the
distributions produced by the anomalous couplings in relation to those
produced by contact terms, as there has been no detailed comparison of these approaches.  
We have tried, in as general a manner as possible, to rewrite the
anomalous TGC occurring in $e^+e^-\rightarrow Z \gamma$ in
terms of contact-type interactions. As it happens, the effective
couplings from the former (anomalous TGC) after reducing to effective couplings with
the $q^2$ dependence of the propagators accounted for, appear
quite different at first sight from the latter (apart from the $q^2$
dependence which is assumed to be absent),  especially since
the anomalous couplings are written down in terms of the Levi-Civita
symbols. At first instance the complete mapping has not been possible because 
in some cases, in the anomalous TGC sector, the basis chosen has been
one that involves the Levi-Civita symbol (CP conserving case).
The conventional treatment of contact interactions does not
involve this symbol.  
However, it is possible through the use
of Dirac matrix identities to choose an equivalent basis for the contact
interactions as well, which could lead to a direct identification.
We have studied the structures in detail
and uncovered these relations so as to establish the correspondence.
We have found that apart from the 
contact interactions studied earlier~\cite{BASDR1, BASDR2, Abraham:1993zh},
a coupling containing three Dirac matrices is also required. 
The form factor containing the three Dirac matrices was introduced in~\cite{Abraham:1998cm}
and the authors have also pointed out that this form factor receives a contribution
from a dimension-8 operator of the form $\bar{l}\gamma^\mu l \epsilon_{\mu\nu\sigma\tau} D^\nu B^{\sigma \lambda} 
B^\tau_\lambda$, which is CP even and was considered earlier in Ref.~\cite{Drell:1984aw}.

In order to make contact with experiment, it is important to ask what 
the contributions of the TGCs at leading order would be 
to the diboson distribution, in the presence of the two kinds of polarization.  
We study this using realistic degrees of polarization, and with the design 
luminosity at the various proposed ILC energies. We limit ourselves to
centre-of-mass energy of 800 GeV along with an integrated luminosity
of 500 fb$^{-1}$. 
Since the BSM contribution from the contact interactions or the 
effective couplings can be measured as deviations from the SM
predictions in various kinematic distributions, we have carried out a 
thorough numerical analyses by the construction of various asymmetries.
In particular the effect of beam polarization has been concentrated upon.
In our previous work~\cite{Ananthanarayan:2011fr, Ananthanarayan:2004eb}, we were concerned 
only with the dimension-six CP-violating operators.  Explicit distributions in the presence
of longitudinal polarization (LP) and transverse polarization (TP)
 were obtained for this case.  However, such an analysis
has not been performed for the dimension-eight CP-violating operator, nor 
for any of the CP-conserving cases, at least
not in the forms discussed in these references.  One of the aims of
this work is to obtain such distributions so as to set the stage for
a thorough comparison with the types of distributions obtained with
the contact interactions.

The layout of the paper is as follows. The process \eegz is
discussed in Sec.~\ref{sec:formalism}, which is divided into three subsections.
We list in Sec.~\ref{subsec:formalism1}
the most general $Z\gamma V^*$ coupling, where $V= Z, \gamma$ and present the 
distributions in the presence of the anomalous couplings with polarized 
beams, both TP and LP. The new physics effect in the form of the contact interactions
will be discussed in Sec.~\ref{subsec:contact_int}
and the mapping of contact interactions with triple gauge boson couplings is addressed in 
Sec.~\ref{subsec:reduction}.
The CPT properties of the different anomalous couplings are discussed
in Sec.~\ref{sec:discrete}. 
In Sec.~\ref{sec:asymmetries} we discuss how angular asymmetries may be
constructed which could be used to get information on the couplings.
We do a full numerical analysis on the anomalous couplings and give limits on those
in Sec.~\ref{sec:results}. Finally we conclude in Sec.~\ref{sec:conclusion}. The 
Appendix~\ref{appendix:reduction} discusses the reduction of the anomalous TGCs with the 
Levi-Civita symbol to an equivalent basis of the contact interactions.

\section{Formalism for the process \eegz }\label{sec:formalism}

In this section we discuss the properties of the process
\begin{equation}
e^+(p_+,s_+)+e^-(p_-,s_-)\rightarrow Z(k_2,h_Z) + \gamma (k_1,h_\gamma),
        \label{process1}
\end{equation}
where $h_\gamma$ can take values $\pm 1$ and the value for $h_Z$
can be $\pm 1$ and 0. 
\begin{figure}[h!]
\centering 
\includegraphics[width=10cm, height=8cm]{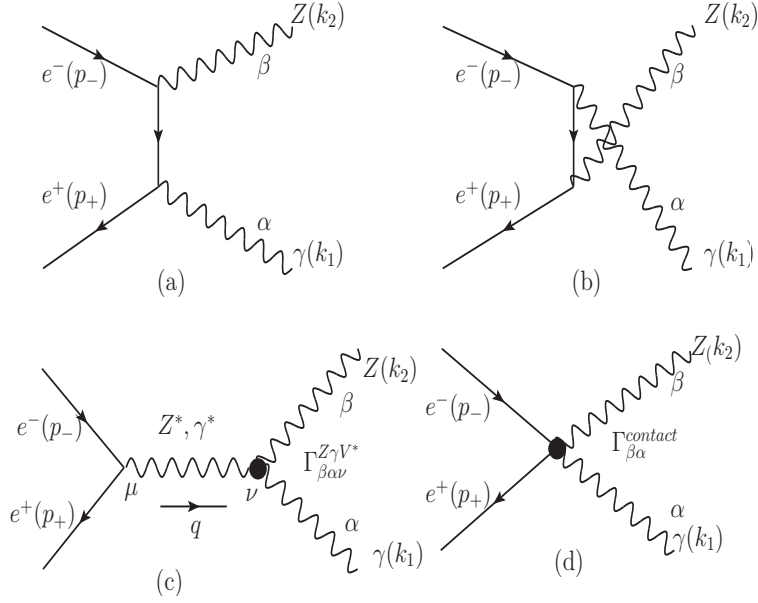}
\caption{Feynman diagrams contributing to the  neutral gauge boson 
production. Diagrams ($a$) and ($b$) are the SM contributions. 
Diagram ($c$) corresponds to contributions from the anomalous $Z\gamma Z$
and $Z\gamma \gamma $ couplings and diagram ($d$) corresponds to contribution 
from the contact interactions.}
\label{process}
\end{figure}
\noindent
In Fig.~\ref{process}, we show the different diagrams which contribute to
neutral gauge boson pair production. The first two diagrams ($a$ and $b$) show the
leading contribution coming from the standard model $t$- and $u$-channel
electron exchanges.
The new-physics effect in the form of anomalous TGCs
due to the $s$-channel $Z$ and $\gamma$ exchanges is shown in the third diagram ($c$),
which will be discussed in detail in Sec.~\ref{subsec:formalism1}.
The effect due to contact interactions is shown in the final diagram ($d$),
and will be the matter of discussion in the upcoming Sec.~\ref{subsec:contact_int}. 
In the final subsection~\ref{subsec:reduction} we 
present a detailed discussion of the TGCs in terms of  the framework of
contact interactions.

\subsection{BSM physics with anomalous triple gauge boson couplings}\label{subsec:formalism1}

The $Z\gamma$ production may have a contribution from the anomalous $Z\gamma Z^*$ or $Z\gamma \gamma^*$
couplings through the $s$ channel, where $Z,~\gamma$
are on shell, while $Z^*/~\gamma^*$ is off shell. Since we neglect the
electron mass, when the off-shell photon or $Z$ couples to fermions, the
corresponding current is conserved.
Assuming $U(1)_{em}$ gauge invariance and Lorentz invariance, 
the most general anomalous $Z\gamma V$ coupling, where $V = Z^*,~\gamma^*$
is given by
\begin{eqnarray}\label{hag_1}
 \Gamma^{Z\gamma Z^*}_{\beta\alpha\nu}(k_2,k_1,q) & = & \frac{e (s-m_Z^2)}{2 m_Z^{2}} 
 \left\{ h_1^Z \left(k_{1\nu}g_{\beta\alpha}-k_{1\beta}g_{\nu\alpha}\right)
 +  \frac{h_2^Z}{m_Z^{2}} q_\beta\left(
 q\cdot k_1 g_{\nu\alpha}-k_{1\nu} q_\alpha\right) \right. \nonumber \\
 &+&  \left. h_3^Z \epsilon_{\nu \beta \alpha \rho}k_{1}^{\rho} 
 +\frac{h_4^Z}{m_Z^{2}} q_\beta \epsilon_{\nu \alpha \rho \sigma} 
	q^\rho k^{1\sigma} \right.\nonumber \\
 &+&\left. \frac{h_5^Z}{2 m_Z^2} \left[(s-m_Z^2)\epsilon_{\alpha \nu \beta \sigma}(k_2+q)^\sigma
-4k_{2\alpha}\epsilon_{\nu \beta \tau \sigma}k_1^\tau q^\sigma
\right]\right\}.
\end{eqnarray}

\begin{eqnarray}\label{hag_2}
 \Gamma^{Z\gamma \gamma^*}_{\beta\alpha\nu}(k_2,k_1,q) & = & \frac{e s}{4 s_W c_W m_Z^{2}} 
 \left\{ h_1^\gamma \left(k_{1\nu}g_{\beta\alpha}-k_{1\beta}g_{\nu\alpha}\right)
 +  \frac{h_2^\gamma}{m_Z^{2}} q_\beta\left(
 q\cdot k_1 g_{\nu\alpha}-k_{1\nu} q_\alpha\right) \right. \nonumber \\
 &+&  \left. h_3^\gamma \epsilon_{\nu \beta \alpha \rho}k_{1}^{\rho} 
 +  \frac{h_4^\gamma}{m_Z^{2}} q_\beta \epsilon_{\nu \alpha \rho \sigma} 
	q^\rho k^{1\sigma} \right\}.
\end{eqnarray}

We note that the coupling $\Gamma^{Z\gamma V^*}_{\beta\alpha\nu}$ was first 
written down in~\cite{Hag}.
However, \cite{Hag} did not have the $h_5^Z$ term. 
The unusual anomalous $Z\gamma Z^*$ vertex in the $h_5^Z$ term, to 
our knowledge, has 
not been noted in the literature. Surprisingly, it does not have a
$Z\gamma\gamma^*$ counterpart.

We have scaled the coupling constants by a factor
of $1/2$ in case of $\Gamma^{Z\gamma Z^*}$ and $1/(4 s_W c_W)$ in case of $\Gamma^{Z\gamma \gamma^*}$,
in relation to those in~\cite{Hag}. This has been done to effect a simple comparison with the 
contact interactions case, where such factors are already absorbed into the definition
of the relevant couplings. 
 The choice is to either rescale the $h_i^V$ terms
of \cite{Hag} or to rescale the contact terms of~\cite{BASDR1, BASDR2}, 
and we choose the former.


The effective Lagrangian generating the vertices of 
Eq.~(\ref{hag_1}) is given by
\begin{eqnarray}\label{lag1}
\mathcal{L}^{Z\gamma Z^*} &=& \frac{e}{2}\left\lbrace \frac{h_1^Z}{m_Z^2}(\partial^\sigma Z_{\sigma\nu})
Z_\alpha F^{\nu \alpha}+
\frac{h_2^Z}{m_Z^4}\left[\partial_\beta  \partial_\alpha
(\Box + m_Z^2) Z_\nu \right] Z^\beta F^{\nu \alpha} \right. \nonumber \\
&+& \left. \frac{h_3^Z}{m_Z^2} (\partial_\sigma Z^{\sigma\rho}) Z^\beta \tilde{F}_{\rho\beta} 
+\frac{h^Z_4}{2m_Z^4}\left[(\Box +m_Z^2)\partial^\sigma 
Z^{\rho\beta}\right] Z_\sigma \tilde{F}_{\rho \beta} \right.  \nonumber \\
 &+& \left. \frac{h^Z_5}{m_Z^4}\left( \partial^\tau F^{\alpha
\lambda}\right)
\tilde Z_{\alpha\beta}
\partial_\tau \partial_\lambda Z^\beta\right\rbrace,
\end{eqnarray}
whereas the Lagrangian generating the vertices of Eq.~(\ref{hag_2}) is given by 
\begin{eqnarray}\label{lag2}
\mathcal{L}^{Z\gamma \gamma^*} &=& \frac{e}{4 s_W c_W} \left\lbrace
\frac{h_1^\gamma}{m_Z^2}(\partial^\sigma F_{\sigma\nu})
Z_\alpha F^{\nu \alpha} + \frac{h_2^\gamma}{m_Z^4}\left[\partial_\beta  \partial_\alpha 
\partial^\rho F_{\rho \nu}\right] Z^\beta F^{\nu \alpha} \right. \nonumber \\
&+& \left. \frac{h_3^\gamma}{m_Z^2}(\partial_\sigma F^{\sigma\rho})Z^\beta \tilde{F}_{\rho\beta} +
\frac{h^\gamma_4}{2m_Z^4}\left[\Box \partial^\sigma F^{\rho\beta}\right] Z_\sigma 
\tilde{F}_{\rho \beta}\right\rbrace.
\end{eqnarray}
Here, 
\beq
F_{\mu\nu} = \partial_\mu A_\nu - \partial_\nu A_\mu; \;\;
Z_{\mu\nu} = \partial_\mu Z_\nu - \partial_\nu Z_\mu
\eeq
and 
\beq
\tilde F_{\mu\nu} = \frac{1}{2} \epsilon_{\mu\nu\alpha\beta} F^{\alpha\beta};\;\;
\tilde Z_{\mu\nu} = \frac{1}{2} \epsilon_{\mu\nu\alpha\beta} Z^{\alpha\beta}.
\eeq

The  matrix element from the SM $t$- and $u$-channel electron
exchanges, and the anomalous coupling introduced by the vertices of 
Eqs.~(\ref{hag_1}) and (\ref{hag_2}), which introduce respectively 
diagrams with
$s$-channel $Z$ and $\gamma$ exchanges, is given by
\begin{equation}
{\cal M}= {\cal M}_1 +{\cal M}_2 +{\cal M}_3 +{\cal M}_4 ,
       \label{amplitude}
\end{equation}
where
\be
\barr{rcl}
{\cal M}_1 & = & \displaystyle
   \frac{e^2}{4c_Ws_W}\, \bar{v}(p_+)\: \sleps(k_2) (g_V - g_A\gamma_5)
   \frac{1}{\slp_- - \slk_1}\: \sleps(k_1)\: u(p_-),
       \\[2ex]
{\cal M}_2 &=& \displaystyle \frac{e^2}{4c_W s_W}\, \bar{v}(p_+) \: \sleps(k_1)
  \frac{1}{\slp_- - \slk_2} \sleps(k_1) (g_V-g_A\gamma_5) u(p_-),
       \\[2ex]
{\cal M}_3 & = & \displaystyle \frac{ie}{2 c_W s_W}\,
       \bar{v}(p_+)\gamma _\mu (g_V-g_A\gamma_5)u(p_-)
   \frac{(-g^{\mu\nu} + q^\mu q^\nu / m_Z^2)}
        {q^2-m_Z^2}
  \Gamma^{Z\gamma Z^*}_{\beta\alpha\nu}(k_2,k_1,q) \epsilon^\alpha(k_1)
        \epsilon^\beta(k_2),
     \\[2ex]
{\cal M}_4 &= & \displaystyle
      ie\,
   \bar{v}(p_+)\gamma _\mu u(p_-)
         \frac{(-g^{\mu\nu})}
              {q^2}
     \Gamma^{Z \gamma \gamma^*}_{\beta\alpha\nu}(k_2,k_1,q)
      \epsilon^\alpha (k_1)\epsilon^\beta(k_2).
\earr
     \label{matrix elem}
\ee
Here, the vector and axial-vector couplings of the $Z$ to the electron 
are given by
\be
g_V =-1 + 4 s_W^2,~~~ g_A = -1,
\ee
where $c_W = \cos\theta_W$, $s_W = \sin\theta_W$, $\theta_W$ being the 
weak mixing angle.

The three-index object  $\Gamma^{Z \gamma V^*}_{\beta\alpha\nu}(k_2,k_1,q)$ is 
effectively contracted with  $-\gamma_\nu/q^2$ in case 
of $\gamma$ and 
$(-\gamma_\nu + \slashed{q} q^\nu / m_Z^2)/ 
(q^2-m_Z^2)$ in case of $Z$ boson, which yields a convenient two-index object which we denote 
as $\Gamma^{Z,\gamma}_{\alpha\beta}$.  We now list in 
Table~\ref{tab:red_form} the various terms in
$\Gamma^{Z,\gamma}_{\alpha\beta}$ in a much simplified form
after dealing with the redundancies, and after dropping $\slashed{q}$
terms which vanish (in the limit of vanishing electron mass) 
on using the Dirac equation.
\begin{table}
\begin{center}
\begin{tabular}{||c|c||}\hline
  $h^V_i$ & $\Gamma^{Z,\gamma}_{\alpha \beta}$ \\ \hline \hline
$h^{\gamma,Z}_1$ &$\gamma^\alpha k_1^\beta-g^{\alpha\beta} \slashed{k}_1
$ \\ 
\hline
$h^{\gamma,Z}_2$  &$k_1^\beta(\slashed{k}_1k_2^\alpha-\gamma^\alpha\frac{s-m_Z^2}{2})$ \\ 
\hline
$h^{\gamma,Z}_3$  &$\gamma _{\nu } \epsilon ^{\alpha \beta \nu k_1}$ \\ \hline
$h^{\gamma,Z}_4$  &$-\gamma_{\nu } k_1^\beta \epsilon ^{\alpha \nu
   k_1 q}$ \\ \hline
$h^Z_5$ & $\gamma_{\nu } \left(2 k_2^\alpha \epsilon
   ^{\beta k_2 \nu  q}+\frac{m_Z^2-s}{2} \left(\epsilon
   ^{\alpha  \beta k_2 \nu }+\epsilon ^{\alpha  \beta  q \nu 
   }\right)\right)$  \\ \hline
\end{tabular}
\caption{The two-index object $\Gamma^{Z,\gamma}_{\alpha\beta}$ 
obtained by contracting the three-index 
object  $\Gamma^{Z \gamma V^*}_{\beta\alpha\nu}(k_2,k_1,q)$  with  
$\frac{(-\gamma_\nu + \slashed{q} q^\nu / m_Z^2)} 
{q^2-m_Z^2}$ in case of the $Z$  and $\frac{-\gamma_\nu}{q^2}$ in case 
of $\gamma$. A factor $(g_V - g_A \gamma_5)$ has to be multiplied on the 
right for all 
the $\Gamma^Z_{\alpha\beta}$ terms. An overall factor of $m_Z^{-2}$ has to be
included for the $h_1^{Z, \gamma}$ and $h_3^{Z, \gamma}$ terms, and a factor 
$m_Z^{-4}$ for the rest.}
\label{tab:red_form}
\end{center}
\end{table}


When the $e^-$ and $e^+$ beams have longitudinal polarizations $P_L$ 
and $\overline{P}_L$, we obtain the differential cross section
for the process (1) to be 
\be
\displaystyle
\frac{d\sigma}{d\Omega    }_L =
    {\cal B}_L\left(1-P_L\overline{P}_L\right)
\left[
	\mathcal{A}_L
       \frac{1}{\sin^2 \theta}
          \left( 1 + \cos^2 \theta + \frac{4 \sbar}{( \sbar - 1)^2}
           \right)
     + C_L              
\right]  \; ,
    \label{diff c.s.L}
\ee
where
\begin{eqnarray}
& \displaystyle
\sbar  \equiv  \frac{s}{m_Z^2},\,\,
   {\cal B}_L  = \frac{\alpha^2}{16 \sin^2\theta_W m_W^2 \sbar}
     \left( 1 - \frac{1}{\sbar}   \right)
     , & 
\end{eqnarray}
with
\begin{eqnarray}
P &=& \frac{P_L - \overline{P}_L}{1- P_L \overline{P}_L},\\
\mathcal{A}_L &=&  (g_V^2+g_A^2-2Pg_Vg_A), \\
C_L &=&\sum_{V=Z, \gamma}\left[\sum_{i=1}^{2} ({\rm Im}~h^V_i) L^V_i 
+ \sum_{i=3}^{4} ({\rm Re}~h^V_i) L^V_i\right]
+ ({\rm Re}~h^Z_5) L^Z_5 . \label{long_Li}
\end{eqnarray}
We choose the convention that
$P_L, \bar{P}_L$ are negative (positive) for left-handed (right-handed)
polarization. 
$C_L$ is a linear combination of the couplings $h^V_{i}$, $(i=1,\dots
,5)$, where
$V = Z,~\gamma$ for $i = 1 - 4$. 
We list in Table~\ref{xytable} the coefficient of each coupling $L^V_i$
in the expression for $C_L$, Eq.~(\ref{long_Li}) against the coupling.
\begin{table}
\begin{center}
\begin{tabular}{||c|c||}\hline
  Coupling & Coefficient \\ \hline \hline
Im $h^Z_1$ &$-\frac{1}{2}\mathcal{A}_L(\bar{s}-1) \cos \theta$ \\ 
Im $h^\gamma_1$ 
&$-\frac{1}{2}(\bar{s}-1)(g_V - P g_A) \cos \theta$ \\ \hline
Im $h^Z_2$  &$\frac{1}{4}\mathcal{A}_L\sbar (\sbar-1) \cos\theta$ \\ 
Im $h^\gamma_2$  &$\frac{1}{4}
     \sbar(\bar{s}-1)(g_V - P g_A ) \cos \theta$ \\ \hline
Re $h^Z_3$ & $-\frac{1}{2}(\bar{s}+1) 
\left(2 g_Vg_A - P(g_V^2+g_A^2)\right)$ \\ 
Re $h^\gamma_3$  &$-\frac{1}{2}
    (\bar{s}+1)(g_A - P g_V)$ \\ \hline
Re $h^Z_4$ &$\frac{1}{4}
\sbar (\sbar-1)\left(2 g_Vg_A - P(g_V^2+g_A^2)\right)$ \\  
Re $h^\gamma_4$  &$\frac{1}{4}\sbar 
(\bar{s}-1)(g_A - P g_V)$ \\ \hline
Re $h^Z_5$ &$-\frac{1}{4}
(1+6 \bar{s}+\bar{s}^2)\left(2 g_Vg_A - P(g_V^2+g_A^2)\right)$ \\ \hline
\end{tabular}
\caption{The coefficients $L_i^V$ of individual new couplings in the
expression for the longitudinal 
polarization dependent part $C_L$, Eq.~(\ref{long_Li}), of the cross section}
\label{xytable}
\end{center}
\end{table}

The differential cross section for transverse polarizations $P_T$ and $\overline{P}_T$ 
of $e^-$ and $e^+$ is given by
\be
\displaystyle
\frac{d\sigma}{d\Omega}_T =
    {\cal B}_T
\left[
	\mathcal{A}_T
       \frac{1}{\sin^2 \theta}
          \left( 1 + \cos^2 \theta + \frac{4 \sbar}{( \sbar - 1)^2}
	+ P_T \overline{P}_T\frac{g_V^2-g_A^2}{g_V^2+g_A^2}
	\sin^2 \theta \cos 2\phi 
           \right)
     + C_T              
\right]  \; ,
    \label{diff c.s.T}
\ee
where $\bar{s}$ is as defined before, 
\begin{eqnarray}
\displaystyle
   {\cal B}_T &=& \frac{\alpha^2}{16 \sin^2\theta_W m_W^2 \sbar}
     \left( 1 - \frac{1}{\sbar}   \right), \\
     C_T &=& \sum_{V=Z, \gamma}\left\lbrace \sum_{i=1}^{4} \left[({\rm Im}~h^V_i) T^{V,I}_i + ({\rm
Re}~h^V_i) T^{V,R}_i \right] \right\rbrace \nonumber \\ 
          &+& ({\rm Im}~h^Z_5) T^{Z,I}_5 + ({\rm Re}~h^Z_5) T^{Z,R}_5
     \label{trans_Ti}
\end{eqnarray}
with $\mathcal{A}_T = (g_V^2 + g_A^2)$.
$C_{T}$ in Eq.~(\ref{trans_Ti}) is a linear combination of the couplings and $V = Z,\gamma$,
and the non-vanishing coefficients $T^{V,I}_i$ and $T^{V,R}_i$ of the various couplings in 
$C_{T}$ are presented in Tables~\ref{trans_table1} and~\ref{trans_table2}.

\begin{table}
\begin{center}
\begin{tabular}{||c|c||}\hline
  Coupling & Coefficient \\ \hline \hline
Im $h^Z_1$ &$-\frac{1}{2}(\bar{s}-1) \cos \theta
(g_V^2 + g_A^2 -(g_V^2 - g_A^2)P_T \overline{P}_T \cos 2\phi)$ \\ 
Im $h^\gamma_1$ 
&$-\frac{1}{2}(\bar{s}-1) g_V \cos \theta
(1 - P_T \overline{P}_T \cos 2\phi )$  \\ \hline
Im $h^Z_2$  &$\frac{1}{4}\sbar (\sbar-1) \cos\theta
(g_V^2 + g_A^2 -(g_V^2 - g_A^2)P_T \overline{P}_T \cos 2\phi)$ \\ 
Im $h^\gamma_2$  &$\frac{1}{4}
\sbar(\bar{s}-1) g_V \cos \theta(1 - P_T \overline{P}_T \cos 2\phi ) $ \\ \hline
Im $h^Z_3$  &$\frac{1}{2}(\sbar-1) 
(g_V^2 - g_A^2)P_T \overline{P}_T \sin 2\phi$ \\ 
Im $h^\gamma_3$ & $\frac{1}{2}(\sbar-1) 
g_V P_T \overline{P}_T \sin 2\phi$ \\ \hline
Im $h^Z_4$  &$-\frac{1}{4}\sbar(\sbar-1) 
(g_V^2 - g_A^2)P_T \overline{P}_T \sin 2\phi$ \\ 
Im $h^\gamma_4$ &$-\frac{1}{4}\sbar(\sbar-1) 
g_V P_T \overline{P}_T \sin 2\phi$ \\ \hline
Im $h^Z_5$  &$\frac{1}{4}(\sbar^2-1) 
(g_V^2 - g_A^2)P_T \overline{P}_T \sin 2\phi$ \\ \hline
\end{tabular}
\caption{The coefficients $T^{V,I}_i$ of the imaginary part of the 
individual new couplings
in the expression for the transverse polarization dependent part $C_T$, 
Eq.~(\ref{trans_Ti}), of the cross section.
Only the non-zero entries are listed here.}
\label{trans_table1}
\end{center}
\end{table}
\begin{table}
\begin{center}
\begin{tabular}{||c|c||}\hline
  Coupling & Coefficient \\ \hline \hline
Re $h^\gamma_1$ 
& $ \frac{1}{2} (\bar{s}-1) g_A \cos \theta P_T \overline{P}_T \sin 2\phi$ \\ \hline
Re $h^\gamma_2$ 
& $-\frac{1}{4} \sbar (\bar{s}-1) g_A                                 
\cos \theta P_T \overline{P}_T \sin 2\phi$ \\ \hline
Re $h^Z_3$ & $-(\bar{s}+1) 
g_A g_V$ \\
Re $h^\gamma_3$  &$-\frac{1}{2}
g_A((\sbar + 1)+ (\bar{s}-1) P_T \overline{P}_T \cos 2 \phi 
)$ \\ \hline
Re $h^Z_4$ &$\frac{1}{2}
\sbar (\sbar-1) g_Vg_A $ \\  
Re $h^\gamma_4$  &$\frac{1}{4}\sbar 
(\bar{s}-1)g_A (1 + P_T \overline{P}_T \cos 2\phi)$ \\ \hline
Re $h^Z_5$ &$-\frac{1}{2}g_A g_V (1+6 \sbar + \sbar^2)$ \\ \hline
\end{tabular}
\caption{The coefficients $T^{V,R}_i$ of the real part of the individual new 
couplings in the expression for the transverse 
polarization dependent part $C_T$, Eq.~(\ref{trans_Ti}), of the cross section.
Only the non-zero entries are listed here.}
\label{trans_table2}
\end{center}
\end{table}
We have kept the anomalous terms up to leading order
since they are expected to be small. 
In the above expressions, $\theta$ is the angle between the photon and $e^-$ direction,
with the $e^-$ direction chosen as the $z$ axis. The azimuthal angle
between the photon and the electron momentum direction is chosen to be 
$\phi$. The transverse polarization
of the electron is chosen along the $x$ axis, whereas the positron 
polarization direction is chosen parallel to the electron  
polarization direction. 

It can be seen from Tables~\ref{xytable},~\ref{trans_table2}
that Re $h^Z_{1,2}$ does not contribute 
to the distribution, with or without beam polarization. 
The question of isolating  Re $h^Z_1$
to leading order was recently addressed by us~\cite{Ananthanarayan:2011fr}, where we pointed 
out that it would be
possible to fingerprint this anomalous coupling if the final-state spins are resolved.
Analogously the contribution of Re $h^Z_{2}$ can be studied 
by analyzing the spin of the final-state particles. 
Tables~\ref{xytable},~\ref{trans_table1} and~\ref{trans_table2} also show that some of the anomalous couplings 
either depend on LP or TP or both, like the anomalous couplings Re $h^\gamma_{1,2}$, 
Im $h^{Z, \gamma}_{3,4}$
only give contributions in the presence of TP. It will therefore be possible to map the correspondence 
between these anomalous form factors and the contact interactions by studying the behaviour of the 
distributions in the presence of different beam polarizations.

In the next subsection, we turn to the issue of parametrizing the BSM physics 
in terms of contact interactions, viz., ones where all the new physics is 
integrated out, and only kinematic information is encoded in the vectors on hand.  
The case of anomalous TGC can be mapped to this, after accounting for the 
(trivial) momentum dependence coming from the propagators. The non-trivial kinematic 
structure due to anomalous TGC would form a proper subset of the general considerations,
which we seek to establish.  The two-index object introduced earlier, provides the 
required bridge to do this.

\subsection{BSM physics in the form of contact interactions}\label{subsec:contact_int}

We now introduce BSM physics arising from contact \eegz interactions as shown 
in the Feynman diagram ($d$) of Fig.~\ref{process}.
The corresponding matrix element for the process of Eq. (\ref{process1}) 
in the presence of contact interactions
will be of the form :
\begin{equation}
{\cal M}= {\cal M}_1 +{\cal M}_2 +{\cal M}_{contact}  ,
       \label{amplitude2}
\end{equation}
where ${\cal M}_{1,2}$ are defined before in Eq.~\ref{matrix elem}, and
\begin{equation}
{\cal M}_{contact} =  \displaystyle
      \frac{ie^2}{4 c_W s_W}\,
   \bar{v}(p_+) \Gamma'_{\alpha\beta} u(p_-)
               \epsilon^\alpha (k_1)\epsilon^\beta(k_2)
\end{equation}
The vertex factor $\Gamma_{\alpha\beta}$ of contact interactions was
studied earlier in~\cite{Abraham:1998cm, BASDR1, BASDR2}, where it was
parametrized in the form :
\begin{equation}
 \Gamma^{contact}_{\alpha\beta}= \frac{ie^2}{4 c_W s_W}\Gamma'_{\alpha\beta}, 
\end{equation}
where
\begin{eqnarray}\label{anomcc}
& \displaystyle \Gamma'_{\alpha\beta}= 
\left\{\frac{1}{m_Z^4}\left((v_1+  a_1 \gamma_5)\gamma_\beta(
2 p{_-}_\alpha (p_+\cdot k_1)-
2 p{_+}_\alpha (p_-\cdot k_1)) + \right. \right. & \nonumber \\ 
& \displaystyle \left. \left.
 ((v_2+  a_2 \gamma_5) p{_-}_\beta + 
(v_3+  a_3 \gamma_5) p{_+}_\beta)(\gamma_\alpha 2 p_-\cdot k_1-
2 p{_-}_\alpha \slk_1)+
\right. \right.& \nonumber \\
& \displaystyle \left. \left.
 ((v_4+  a_4 \gamma_5) p{_-}_\beta +
(v_5+  a_5 \gamma_5) p{_+}_\beta)(\gamma_\alpha 2 p_+\cdot k_1-
2 p{_+}_\alpha \slk_1)\right)+ \right. & \nonumber \\
& \displaystyle \left. \frac{1}{m_Z^2} \left(
(v_6+  a_6 \gamma_5)(\gamma_\alpha k_{1\beta}-\slk_1 
g_{\alpha\beta}) + (v_7+  a_7 \gamma_5) \slashed{k}_1 \gamma_\alpha \gamma_\beta \right)
 \right\} .&
\end{eqnarray}
The above is the most general form consistent with Lorentz and gauge
invariance, and written in terms of an odd number of $\gamma$ matrices,
so that chirality is conserved by the vertex. 

When the only  BSM interactions present are the triple-gauge boson
couplings shown in Eqs. (\ref{hag_1}) and (\ref{hag_2}), the vertex factor $\Gamma'_{\alpha\beta}$ is 
effectively the sum of the $\Gamma^{Z,\gamma}_{\alpha \beta}$ terms of Table 
\ref{tab:red_form} appropriately weighted:
\be \label{contact-triple}
\Gamma'_{\alpha\beta} = \sum_{i=1,3} m_Z^{-2} [h_i^\gamma
\Gamma_{\alpha\beta}^{\gamma,i}+ h_i^Z \Gamma_{\alpha\beta}^{Z,i}(g_V-\gamma_5
g_A)] + \sum_{i=2,4,5} m_Z^{-4} [h_i^\gamma \Gamma_{\alpha\beta}^{\gamma,i}+
h_i^Z \Gamma_{\alpha\beta}^{Z,i}(g_V - \gamma_5 g_A)]
\ee
Of course, it is always possible that there are further interactions
present which do not contribute to the triple-gauge couplings, but
contribute in the form of contact interactions.
One of our aims here is to make a correspondence between the form
factors $v_i$, $a_i$ written in the contact interactions and those in
the triple-gauge boson vertices.
The distributions
arising from the new couplings (with the exception of $v_7$ and $a_7$) in the presence of both longitudinal
and transverse polarization were given in~\cite{BASDR1, BASDR2}. 
We would also like to compare these distributions with those obtained
in the previous section. 

The contributions of the new contact interactions to the 
the cross section with longitudinal and
transverse polarizations of the beams,  as defined respectively 
by $C_L$ and $C_T$ of Eqs. (\ref{diff c.s.L}) and (\ref{diff c.s.T}),
are given by
\begin{eqnarray}
& \displaystyle
C_{L}  =  
        \frac{1}{4}
    \left\{\sum_{i=1}^7
\left(	(g_V-Pg_A) {\rm Im}v_i+ (g_A-Pg_V) {\rm Im}a_i\right) X_i 
\right\},&
\end{eqnarray}
and
 \begin{eqnarray}
C_{T}  &=&  \frac{1}{4}
    \left\{\sum_{i=1}^7
        (g_V {\rm Im}v_i+ g_A {\rm Im}a_i) X_i + P_T \overline{P}_T\, \right.  \nonumber \\
        && \left. \times
\sum_{i=1}^7\left(
        (g_V {\rm Im}v_i- g_A {\rm Im}a_i)
    \cos 2\phi+
        (g_A {\rm Re}v_i- g_V {\rm Re}a_i)
                 \sin 2\phi\right)Y_i  \right\}. 
   \label{notation}
\end{eqnarray}
$X_i$ and $Y_i$ ($i = 1, \cdots , 7$) 
are listed in Table~\ref{contact:table}.
\begin{table}\label{contact:table}
\begin{center}
\begin{tabular}{||c|c|c||}\hline
$i$ &  $X_i$ & $Y_i$ \\ \hline \hline
$1$ & $ 2 \sbar (\sbar+1)$ & 0  \\ \hline
$2$ & $  -\sbar (\sbar-1) (\cos\theta-1)$ & $0$\\ \hline
$3$ & $  0 $ & $\sbar (\sbar-1)(\cos\theta-1)$ \\ \hline
$4$ & $  0 $ & $ \sbar (\sbar-1)(\cos\theta+1)$ \\ \hline
$5$ & $ -\sbar (\sbar-1) (\cos\theta+1)$ & $0$ \\ \hline
$6$ & $ -2 (\sbar-1) \cos\theta $ & $ 2 (\sbar-1)\cos\theta$\\ \hline
$7$ & $2 (\sbar - 1) (1 + \cos \theta) + 4$ & $-2 (\sbar - 1) (1 + \cos \theta)$ \\ \hline 
\end{tabular}
\caption{The contribution of the new couplings to the polarization
independent and dependent parts of the cross section.}
\end{center}
\end{table}

In case of the contact interactions it is seen that, with the exception of $v_{6,7}$
and $a_{6,7}$, the anomalous form factors either contribute to the transverse polarization
dependent part, or to the longitudinal polarization dependent and polarization independent
parts of the differential cross section, but not both. The anomalous form factors $v_{6,7}$ and
$a_{6,7}$, on the other hand, contribute to both. 

\subsection{Reduction of anomalous TGC interactions to contact type interactions}\label{subsec:reduction}

In order to make a correspondence between the two approaches, we
compare the matrix elements of Eq. (\ref{matrix elem}) and Eq.
(\ref{amplitude2}), using Eq. (\ref{contact-triple}) and using the forms of
$\Gamma^{Z, \gamma}_{\alpha\beta}$ with the Levi-Civita tensor, if any,
rewritten using the results of the Appendix~\ref{appendix:reduction}. On equating coefficients of
the independent $\gamma$-matrix and tensor combinations, we get the
relations
\begin{eqnarray}
v_1 & = & -2 i  h_Z^5 g_A,  \\ \label{examp1}
v_2 + v_3 + v_4 + v_5 &=& - 2  (h_2^\gamma + h^Z_2  g_V),\\ 
 v_2 + v_3 - v_4 - v_5 &=&  2 i h^Z_4 g_A,
\\
v_2 - v_3 - v_4 + v_5 &=& 0,
\\ 
v_2 - v_3 + v_4 - v_5 &=& 4 i h^Z_5 g_A ,
\\ \label{examp_v6}
v_6 &=& h^\gamma_1 + h^Z_1 g_V - i h^Z_3 g_A + i h^Z_5 g_A (s -
m_Z^2)/(2 m_Z^2), 
\\ \label{examp_v7}
v_7 &=& i (-h^Z_3 g_A + h^Z_5 g_A (s-m_Z^2)/(2 m_Z^2) ),
\end{eqnarray}
and
\begin{eqnarray}
a_1 &=& -  2 i h^Z_5 g_V,
\\
a_2 + a_3 + a_4 + a_5 &=& - 2 h^Z_2  g_A,
\\
 a_2 + a_3 - a_4 - a_5 &=&2  i (h^Z_4 g_V + h^\gamma_4),
\\
a_2 - a_3 - a_4 + a_5 &=& 0,
\\
a_2 - a_3 + a_4 - a_5 &=& 4 i h^Z_5 g_V, 
\\ \label{examp_a6}
a_6 &=& (h^Z_1 g_A - i (h^\gamma_3 + h^Z_3 g_V) + i h^Z_5 g_V
(s-m_Z^2)/(2 m_Z^2) ),
\\ \label{examp_a7}
a_7 &=& i (-h^Z_3 g_V - h^\gamma_3 + h^Z_5 g_V (s-m_Z^2)/(2 m_Z^2) ).
\end{eqnarray}
These may be solved for $v_i$, $a_i$ in terms of the $h_i^V$.
The above relations hold at the amplitude level. In turn,
the distributions generated
by the $v_i$, $a_i$ of the contact interactions
would be indistinguishable from the 
distribution generated by the TGCs with coefficients obeying these
equations. The TGCs being less in number than
the contact interactions, when the contact interactions come from TGCs,
they obey constraints among themselves. These constraints can
then be a test of whether the TGCs describe the full new physics or not.

\section{Discrete symmetries of the BSM interactions}\label{sec:discrete}

In order to study the properties of the different TGCs,
by the construction of different asymmetries, we need to
first understand the CP properties of various terms in the differential
cross section. 
For completeness, we now provide a brief recapitulation of 
the discussion provided 
in the case of contact interactions \cite{BASDR1,BASDR2}, 
which we now extend in the case of
anomalous TGCs. 
Firstly, we consider the case of TP, for which
we note the following relations:
\beq\label{ctheta}
\vec{P}\cdot \vec{k}_1 =\frac{\sqrt{s}}{2} \vert \vec{k}_1 \vert \cos\theta\;,
\eeq
\beq\label{s2ts2p}
(\vec{P} \times \vec{s}_- \cdot \vec{k}_1)( \vec{s}_+\cdot \vec{k}_1) + 
(\vec{P} \times \vec{s}_+ \cdot \vec{k}_1) (\vec{s}_-\cdot \vec{k}_1) 
= \frac{\sqrt{s}}{2} \vert \vec{k}_1 \vert^2 \sin^2\theta \sin 2\phi\; ,
\eeq
\beq\label{s2tc2p}
(\vec{s}_- \cdot \vec{s}_+) (\vec{P}\cdot \vec{P} \vec{k}_1 \cdot \vec{k}_1 - 
\vec{P}\cdot\vec{k}_1 \vec{P}\cdot\vec{k}_1) - 2 (\vec{P}\cdot \vec{P}) 
( \vec{s}_-
\cdot \vec{k}_1) ( \vec{s}_+\cdot \vec{k_1})
 = -\frac{s}{4} \vert \vec{k}_1 \vert^2 
\sin^2\theta \cos 2\phi\; .
\eeq 
In the above equations, $\vec{P}=\frac{1}{2}(\vec{p}_- - \vec{p}_+)$, 
where $p_-$ is the momentum of the electron, and $p_+$ is the 
momentum of the positron. Moreover it is assumed that
$\vec{s}_+=\vec{s}_-$; taking $\vec{s}_+= - \vec{s}_-$ would only give
an overall negative sign for all the terms.
Observing that the vector $\vec{P}$ is C and P odd, that the photon
momentum $\vec{k}_1$ is C even but P odd, and that the spin vectors
$\vec{s}_{\pm}$ are P even, and go into each other under C, 
we can immediately check that only the left-hand side (lhs)
of Eq. (\ref{ctheta}) is CP odd, while the lhs of 
Eqs. (\ref{s2ts2p}) and (\ref{s2tc2p})
are CP even. Of all the above, only the lhs of (\ref{s2ts2p}) is odd under
naive time reversal T.  

Many of these features can be explicitly checked from 
Tables~\ref{trans_table1},~\ref{trans_table2}: we see that the term $\cos \theta$ is
accompanied by the CP violating couplings $h^Z_1,~h^Z_2,~h^\gamma_1,
h^\gamma_2$, whereas the CP conserving couplings $h^Z_3,~h^Z_4,~h^Z_5,
h^\gamma_3,~h^\gamma_4,~h^\gamma_5$ have no $\cos \theta$ dependence. 
It is known that invariance under CPT implies 
that terms with the right-hand side (rhs) of (\ref{ctheta}) by itself, 
or multiplying the rhs of Eq.~(\ref{s2tc2p}) would occur with absorptive 
(imaginary) parts of the form factors, whereas the rhs of Eq.~(\ref{ctheta}) 
multiplied by the rhs of Eq.~(\ref{s2ts2p}) would appear with dispersive (real) 
parts of the form factors. Therefore the imaginary part of the CP-odd terms
always come with a factor of $\cos \theta$ or $\cos \theta \cos 2\phi$
and the real parts are accompanied by the factor $\cos \theta \sin 2\phi$.
Similarly the imaginary part of the CP-even terms, which has no $\cos \theta$
dependence always come with a factor of $\sin 2\phi$
and the real parts are either accompanied with the factor $\cos 2\phi$ or no $\theta, \phi$ dependence.
The CPT dependence of the different anomalous couplings are used to construct the different
asymmetries to be proposed and discussed in the next section.
 
As discussed in the earlier work~\cite{BASDR1, BASDR2},
in case of the contact interactions (Sec.~\ref{subsec:contact_int}),
the coefficients of the combinations of couplings $r_2+r_5$,  $r_3+r_4$,
and of the coupling $ r_6$, ($r_i=v_i, a_i$) have a pure $\cos\theta$
dependence and are CP odd.  On the other hand, the coefficients of $r_1$
and of the remaining linearly independent
combinations $r_2-r_5, \, r_3-r_4, \, (r_i=v_i,a_i)$   
have no $\cos\theta$ dependence. These combinations have been isolated by considering the
tensors accompanying the coefficients $r_i$.  
Keeping in mind the fact that under 
C $p_+ \leftrightarrow p_-$ and $k_{1,2} \leftrightarrow k_{1,2}$,
 these properties may be readily inferred from the form of the tensors
for $i= 1,\dots,6$.
An analysis with the inclusion of $r_7$ is more complicated.  By construction, 
the $r_7$ term has no straightforward transformation property under C.  
An analysis must include $r_6$ and $r_7$ jointly.
Writing the $r_6$ and $r_7$ terms as
$ r_6 \mathcal{O}_6 +r_7 \mathcal{O}_7$,   where $\mathcal{O}_6$ and 
$\mathcal{O}_7$ are Dirac operators
sandwiched between spinors, we can rewrite these terms as
$$(r_6 - r_7) \mathcal{O}_6 +r_7 (\mathcal{O}_6 + \mathcal{O}_7)$$
It may be verified $(\mathcal{O}_6+\mathcal{O}_7)\equiv 
(\gamma_\alpha k_{1\beta}-\slk_1 g_{\alpha\beta} + \slashed{k}_1 \gamma_\alpha 
\gamma_\beta)$ 
is CP even. 
We conclude that while $r_6$ accompanies a purely CP-odd operator
$\mathcal{O}_6$, the
operator multiplying $r_7$ is in part CP odd (viz., $-\mathcal{O}_6$),
and in part CP even ($\mathcal{O}_6+\mathcal{O}_7$).
Thus Eqs.~(\ref{examp_v6}) and (\ref{examp_v7}) and the corresponding
Eqs.~(\ref{examp_a6}) and (\ref{examp_a7}) for the $a$'s are consistent with this,
since $h^{Z,\gamma}_3$ and $h^Z_5$ which are CP even contribute equally to $r_6$ and $r_7$.
This completes our discussion of the discrete symmetry
properties of the BSM physics in the process.

In case of longitudinal polarization, apart from Eq.~(\ref{ctheta}), there is 
another CP-odd quantity, viz., 
\beq\label{cthetapol}
\frac{1}{2}\left(\vec{s}_- + \vec{s}_+\right) \cdot \vec{k}_1 
= \vert \vec{k}_1 \vert \cos\theta\;.
\eeq
While this is also proportional to $\cos\theta$ like (\ref{ctheta}), it is 
expected to appear with a factor $(P_L-\overline{P}_L)$ multiplying it. It is 
also CPT odd, and would therefore occur with the absorptive parts of form 
factors. With all these considerations in view, we now embark on the task of constructing 
suitable asymmetries to isolate the anomalous TGCs which is the aim of the next section.

\section{Angular asymmetries for anomalous TGCs}\label{sec:asymmetries}
In earlier studies, several asymmetries were considered to isolate the effects of contact interactions.
Since, in this work we do not extend that sector, except for the
couplings $v_7$ and $a_7$, we will be primarily
concerned with the task of isolating the anomalous TGCs, which form the 
main focus of our study.
Contact interactions have been brought in for making a correspondence 
and showing that TGCs do not exhaust all possibilities.
The angular distributions defined in Tables~\ref{xytable},~\ref{trans_table1},~\ref{trans_table2}, 
involve several different functions of $\theta$ and $\phi$, such as $\sin 2\phi$,
$\sin 2\phi \cos \theta$, $\sin 2\phi \sin \theta$, $\cos 2\phi$,
$\cos 2\phi \cos \theta$ etc. We next formulate different angular
asymmetries which can be used to determine or disentangle the different linear combinations 
of the anomalous couplings. For all our calculations we have assumed a 
cut-off $\theta_0$ on the polar angle $\theta$ of the photon in the forward and backward
directions in order to stay away from the beam pipe. This cut-off may be chosen to optimize
the sensitivity of the measurement. 

We first present the case of transverse polarization
where we have considered both CP-odd and CP-even asymmetries so as to
determine the anomalous couplings. 
The asymmetries defined in general
are an appropriate asymmetry in $\phi$, $A_{i2}, i=1,2,3,4$, and the
same $\phi$ asymmetry combined with a forward-backward asymmetry
in $A_{i1}, i=1,2,3$. The forward-backward
asymmetry in $A_{i1}$ isolates terms with a $\theta$ dependence of
$\cos\theta$, i.e., it is a CP-odd asymmetry, whereas $A_{i2}$ isolates $\theta$ dependence which is
either trivial, or proportional to $\sin\theta$. The asymmetry $A_{i2}$
is sensitive to the CP-even couplings.
The CP-odd asymmetries are defined as follows\footnote{In case of contact interactions, $A_{11}$
is proportional to (Re $r_6$ - Re $r_7$), and $A_{21,31}$ are proportional to (Im $r_6$ - Im $r_7$). We list 
them here as the coupling $r_7$ was not discussed in~\cite{BASDR1, BASDR2}}:
\begin{eqnarray} 
\displaystyle A_{11}(\thmin)&=&
{1\over \sigma^{SM}}
\sum_{n=0}^3 (-1)^n
\left(
\int_{-\cos \theta_0}^{0} d \cos\theta
 -
\int_{0}^{\cos \theta_0} d \cos\theta \right) 
 \int_{\pi n/ 2}^{\pi(n+1)/  2} d\phi \,
{d \sigma \over d \Omega},   \\
\displaystyle A_{21}(\thmin) & = & 
{1\over \sigma^{SM}}
\sum_{n=0}^3(-1)^n \left(
\int_{-\cos \theta_0}^0 d \cos\theta -
\int_{0}^{\cos \theta_0} d \cos\theta \right)
 \int_{\pi (2 n-1)/4}^{\pi(2 n+1)/4} d\phi \,
{d \sigma \over  d \Omega},   \\
\displaystyle A_{31}(\thmin)&=&\frac{1}{\sigma^{SM}}
\left(\int_{-\cos \theta_0}^{0} d \cos\theta -
\int^{\cos \theta_0}_0 d \cos\theta\right)\left(\int_{-\frac{\pi}{4}}^{\frac{\pi}{4}}
d\phi \frac{d\sigma}{d\Omega}+\int_{\frac{3\pi}{4}}^{\frac{5\pi}{4}}d\phi \frac{d\sigma}{d\Omega}\right),  
\end{eqnarray}
whereas the CP-even asymmetries are\footnote{In case of contact interactions, $A_{12}$
is proportional to  Re $r_7$, and $A_{22,32}$ are proportional to Im $r_7$. We list 
them here as the coupling $r_7$ was not discussed in~\cite{BASDR1, BASDR2}}
\begin{eqnarray}
\displaystyle A_{12}(\thmin)&=&
{1\over \sigma^{SM}}
\sum_{n=0}^3 (-1)^n
\int_{-\cos \theta_0}^{\cos \theta_0} d \cos\theta
  \int_{\pi n/ 2}^{\pi(n+1)/  2} d\phi \,
{d \sigma \over d \Omega} ,  \\
\displaystyle A_{22}(\thmin) &=&
{1\over \sigma^{SM}}
\sum_{n=0}^3(-1)^n 
\int_{-\cos \theta_0}^{\cos \theta_0} d \cos\theta 
 \int_{\pi (2 n-1)/4}^{\pi(2 n+1)/4} d\phi \,
{d \sigma \over  d \Omega} ,  \\
\displaystyle A_{32}(\thmin)&=&\frac{1}{\sigma^{SM}}
\int_{-\cos \theta_0}^{\cos \theta_0} \left(\int_{-\frac{\pi}{4}}^{\frac{\pi}{4}}
d\phi \frac{d\sigma}{d\Omega}+\int_{\frac{3\pi}{4}}^{\frac{5\pi}{4}}d\phi \frac{d\sigma}{d\Omega}\right) ,
\end{eqnarray}
with
\begin{eqnarray}
& \displaystyle \sigma^{SM} \equiv \sigma^{SM}(\thmin)=
\int_{-\cos \theta_0}^{\cos \theta_0} d \cos\theta
\int_{0}^{2 \pi} d\phi \,
{d \sigma_{SM} \over d \Omega}\, . &
\end{eqnarray}
\noindent 
The choice of the asymmetries is such that each  
asymmetry is dependent on a particular form of angular dependence.
For instance in the asymmetry $A_{12}$, only the terms proportional to
$\sin 2\phi$ or $\sin 2\phi \sin\theta$ survive, whereas in case of
$A_{11}$  it is the $\sin 2\phi \cos \theta$ terms which survive. 
The terms proportional to $\sin 2\phi$ or $\sin 2\phi \sin\theta$
are CPT odd and appear with the imaginary part of the 
anomalous couplings whereas the $\sin 2\phi \cos \theta$ terms
are CPT even and appear with the real part of the anomalous couplings,
as discussed in Sec.~\ref{sec:discrete}.
The SM contribution to $A_{11,12}$ is zero, since, as
can be seen from Eq.~(\ref{diff c.s.T}), it has no  $\sin 2\phi$ terms. 
Therefore the  observation of either of these asymmetries at the ILC
will point towards contribution from anomalous couplings.
Similarly $A_{22}$ has terms proportional to $\cos 2\phi$ and
$\cos 2\phi \sin \theta$ and $A_{21}$ has $\cos 2\phi \cos \theta$
dependence. It can be argued like before that the SM contribution to
$A_{21}$ will be zero and $A_{22}$ will occur with the real
parts of the anomalous couplings whereas $A_{21}$ will receive
contribution from the imaginary parts. It can be checked that the other asymmetries
$A_{31,32}$ contain terms which are not proportional to the transverse polarization.

We present below the dependence of the asymmetries on the various 
anomalous couplings.
The CP-odd asymmetries are given by
\begin{eqnarray}
\displaystyle A_{11}(\thmin)&=& \mathcal{B'}_T
g_A P_T \bar{P}_T (\sbar-1) \cos ^2 \theta_0
   (\sbar \rm{Re}h^\gamma_2 -2 {\rm Re}h^\gamma_1),\label{diff_asym_cpodd11} \\
\displaystyle A_{21}(\thmin)&=&  \mathcal{B'}_T P_T \bar{P}_T (-1+\sbar)  
\cos ^2 \theta_0 \left(2 {\rm Im} h^Z_1 (g_V^2-g_A^2) + 2 {\rm Im} h^\gamma_1 g_V \right. \nonumber \\ 
&+& \left. 
 ({\rm Im} h^Z_2 (g_A^2 - g_V^2) - g_V {\rm Im} h^\gamma_2)\sbar \right), \label{diff_asym_cpodd21}\\   
\displaystyle A_{31}(\thmin)&=& \frac{\mathcal{B'}_T}{4}(\sbar-1) \cos ^2 \theta_0 \left\lbrace -g_V (\pi - 2 P_T \bar{P}_T)
(2 {\rm Im} h^\gamma_1 - \sbar {\rm Im} h^\gamma_2) \right. \nonumber \\
 &-& \left. \left[g_V^2(\pi -  2 P_T \bar{P}_T) + g_A^2 (\pi +  2 P_T \bar{P}_T)\right]
(2 {\rm Im} h^Z_1 - \sbar {\rm Im} h^Z_2)\right\rbrace \label{diff_asym_cpodd31},
\end{eqnarray}
and the CP-even asymmetries by
\begin{eqnarray}
\displaystyle A_{12}(\thmin)&=&  2 \mathcal{B'}_T P_T \bar{P}_T  (\sbar-1) 
\left(g_V(-2 {\rm Im}h^\gamma_3 + \sbar {\rm Im}h^\gamma_4) \right. \nonumber \\
 &+& \left. ({\rm Im}h^Z_5 (\sbar+1)+2 {\rm Im}h^Z_3-\sbar {\rm Im}h^Z_4 )(g_A^2-g_V^2)\right) 
\cos \theta_0, \label{diff_asym_cpeven12}\\
\displaystyle A_{22}(\thmin)&=& 2 \mathcal{B'}_T P_T \bar{P}_T \cos \theta_0
 \left(4 (g_A^2-g_V^2) + g_A (\sbar-1)  ( \sbar {\rm Re} h^\gamma_4 - 2
{\rm Re} h^\gamma_3)\right),\label{diff_asym_cpeven22}  \\
\displaystyle A_{32}(\thmin)&=&  -\frac{\mathcal{B'}_T}{2}\left[ \left\lbrace 4 
(g_V^2(\pi +  2 P_T \bar{P}_T) + g_A^2 (\pi -  2 P_T \bar{P}_T))+
g_A \right. \right. \nonumber \\
&& \left. \left.
\left[-2 P_T \bar{P}_T (\sbar-1)(\sbar {\rm Re} h^\gamma_4-2 {\rm Re} h^\gamma_3)
+\pi \left(-\sbar(\sbar-1) {\rm Re} h^\gamma_4 + 2 (\sbar+1) {\rm Re} h^\gamma_3 \right. \right. \right. \right. \nonumber \\
&+& \left. \left. \left. \left. 2 g_V \left[{\rm Re} h^Z_5+2 {\rm Re} h^Z_3 (\sbar+1)+\sbar\left\lbrace
{\rm Re} h^Z_4 (1-\sbar)+{\rm Re} h^Z_5(6+\sbar)\right\rbrace\right]\right)\right]\right\rbrace \cos\theta_0 \right.\nonumber \\ 
&+& \left.  4 \pi \mathcal{A}_T   \frac{1+\sbar^2}{(\sbar-1)^2} 
\log \left(\frac{1- \cos \theta_0}{1+\cos \theta_0}\right)\right]\label{diff_asym_cpeven32},
\end{eqnarray}
where $\mathcal{B'}_T = \mathcal{B}_T/\sigma^{SM}(\theta_0)$, and
\begin{eqnarray}\label{sm_tot}
\sigma^{SM}(\theta_0) = 4 \pi \mathcal{A}_T \mathcal{B}_T  \left[\frac{1+\sbar^2}{(\sbar-1)^2}
\log \left(\frac{1+ \cos \theta_0}{1-\cos \theta_0}\right)  - \cos \theta_0  \right].
\end{eqnarray}

We have also considered a CP-odd asymmetry in the presence of longitudinal
polarization, which is proportional to $\cos\theta$. 
It is shown in Sec.~\ref{sec:discrete}, Eqs.~(\ref{ctheta}),~(\ref{cthetapol}) that the term proportional to
$\cos \theta$ is CPT odd and would therefore always occur with the imaginary part of the 
anomalous couplings.  The asymmetry is a forward-backward asymmetry
\begin{equation}
\displaystyle A^{LP}(\thmin) = \frac{1}{\sigma^{SM}_{LP}}
\left(\int_{-\cos \theta_0}^{0} d \cos\theta-
\int^{\cos \theta_0}_0 d \cos\theta\right)
\int_{0}^{2 \pi} d\phi \,
{d \sigma \over d \Omega} ,
\end{equation}
with the form 
\begin{eqnarray}\label{long_asym}
A^{LP}(\thmin) &=& \frac{\mathcal{B'}_L \pi}{2} (\sbar-1) \cos^2 \theta_0
 \left(-2 \left[(g_V - P g_A) {\rm Im} h^\gamma_1 + \mathcal{A}_L {\rm Im} h^Z_1 \right] \right. \nonumber \\
&+& \left. \left[(g_V - P g_A) {\rm Im}h^\gamma_2 + \mathcal{A}_L {\rm Im} h^Z_2 \right] \sbar \right),
\end{eqnarray}
where $\mathcal{B'}_L = \mathcal{B}_L/\sigma^{SM}_{LP}(\theta_0)$.
In the presence of longitudinal polarization, $\mathcal{B}_T$ is replaced by
$\mathcal{B}_L (1-P_L \bar{P}_L)$ and $\mathcal{A}_T$ is replaced by
$\mathcal{A}_L$ in Eq.~(\ref{sm_tot}) for $\sigma^{SM}_{LP}(\theta_0)$.

In the next section we evaluate these asymmetries numerically and
investigate what limits on couplings may be expected by an experimental
study of the asymmetries.
 
\section{Numerical analysis}\label{sec:results}
The asymmetries listed above receive contributions from combinations of the couplings.
Since the 
number of different types of terms in the angular
distribution is not large, it will not be
possible to disentangle the effects of all the anomalous couplings,
without a full-fledged fit to the distributions. The presence
of all of them at the same time will make the numerical analysis complicated.
We have therefore estimated possible 90\% CL limits on the couplings assuming only one coupling
to be non-zero at a time. For our discussion we have assumed $\sqrt{s}$ of 800 GeV,
along with an integrated luminosity of 500 fb$^{-1}$. The magnitudes of electron and positron
polarization are taken to be 0.8 and 0.6 respectively. When an asymmetry
arises only in the presence of BSM the 90\% CL
limits on the coupling, denoted
by $\mathcal{C}_{lim}$, is related to the value $A$ of the generic
asymmetry for unit value of the 
anomalous coupling by 
\begin{equation}\label{eq:limit}
 \mathcal{C}_{lim} \equiv \frac{1.64}{|A| \sqrt{N_{SM}}},
 \end{equation}
where $N_{SM}$ is the number of SM events. The
coefficient 1.64 may be obtained from statistical tables for hypothesis
testing with one estimator; 
see, e.g., Table 36.1 of Ref.~\cite{Beringer:1900zz}.

We present here our results 
for the best limits obtainable on the anomalous couplings from various asymmetries.
Since the anomalous couplings with $\sin 2\phi$ dependence give non-zero contribution
for the asymmetries $A_{11,12}$, we present our results for this case. Along with it
we also consider the asymmetries $A^{LP}, A_{31}, A_{32}$.
\begin{figure}
  \centering
  \includegraphics[width=7cm, height=5.5cm]{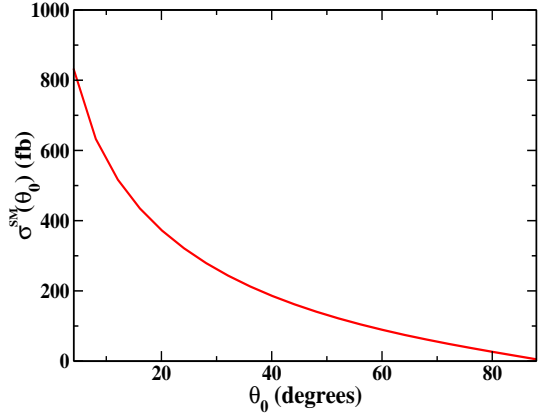}
  \caption{The SM cross section $\sigma^{SM} (\theta_0)$  defined in Eq.~(\ref{sm_tot}) as a function
of the cut-off angle $\theta_0$ at $\sqrt{s}$ = 800 GeV.}
  \label{fig:sm_dist}
\end{figure}
\noindent 
We show in Fig.~\ref{fig:sm_dist} the SM cross section, with a cut-off $\theta_0$
in the forward and backward directions, as a function of $\theta_0$.

\begin{figure}
\vspace{0.4cm}
\begin{minipage}{0.45\linewidth}
  \centering
  \includegraphics[width=6.5cm, height=5cm]{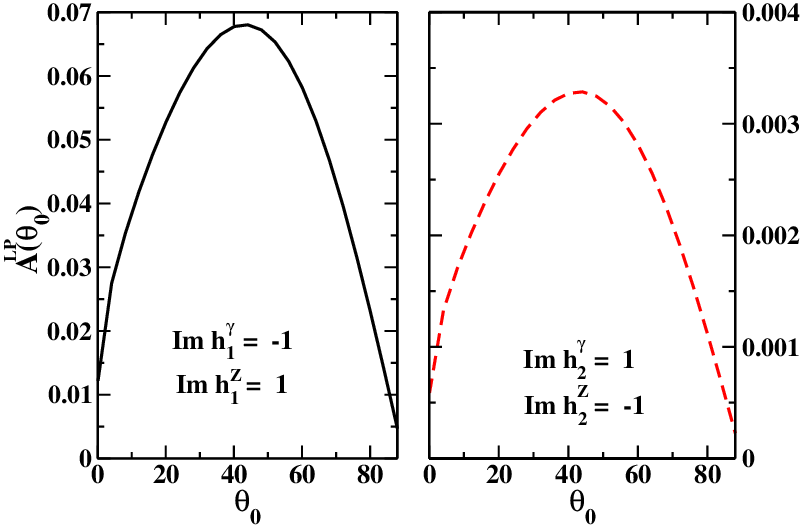}
  \caption{The asymmetry  $A^{LP} (\thmin)$ defined in Eq.~(\ref{long_asym}) as a function
of the cut-off angle $\theta_0$ for the various couplings.}
  \label{fig:long_asym}
\end{minipage}%
\hspace{0.5cm}
\begin{minipage}{0.45\linewidth}
  \centering
  \includegraphics[width=6.5cm, height=5cm]{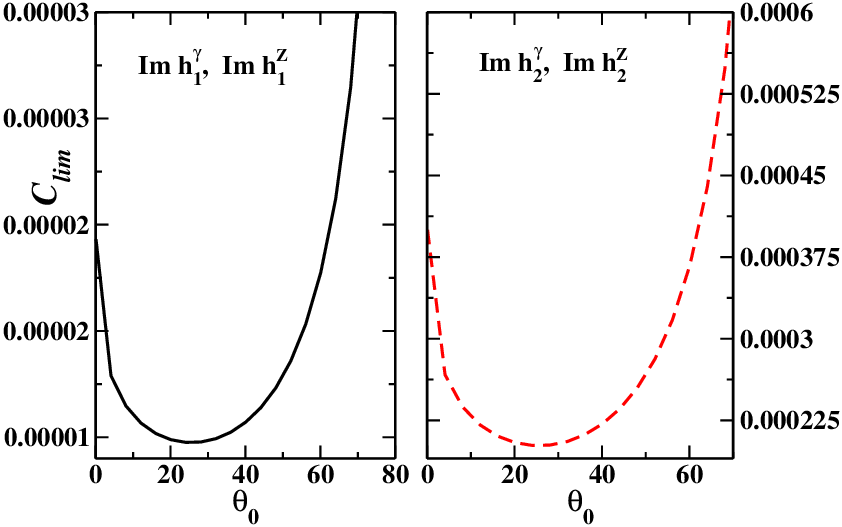}
  \caption{The 90\% CL limits, Eq.~(\ref{eq:limit}) on the anomalous couplings from $A^{LP} (\thmin)$, as a function
of the cut-off angle $\theta_0$.}
  \label{fig:long_limit}
\end{minipage}
\end{figure}

In  case of the longitudinal polarization, we have considered the 
forward-backward asymmetry. It can be seen from Eq.~(\ref{long_asym}) that the SM contribution is
equal to zero and the couplings
which contribute are  Im $h^\gamma_{1,2}$ and Im $h^Z_{1,2}$. The coefficient 
of Im $h^\gamma_1$ and Im $h^\gamma_2$ are dependent
on the choice of beam polarization. For our choice of beam polarization, $P_L = -0.8$
and $\bar{P}_L = 0.6$, the coefficients $(g_V - P g_A)$ and $\mathcal{A}_L$ are 
almost the same apart from a minus sign. Therefore the behaviour of |Im $h^\gamma_1$| 
and |Im $h^Z_1$| will be the same. 
They will however behave differently with unpolarized beams but with less sensitivity.
Fig.~\ref{fig:long_asym} shows the asymmetry $A^{LP} (\thmin)$ as a function
of the cut-off angle $\theta_0$, with the assumption of only one anomalous coupling
being present at a time. We have next shown in Fig.~\ref{fig:long_limit} the 
90\% CL limits that can be obtained on these couplings from the asymmetry.
It can be seen from Fig.~\ref{fig:long_limit} that the limit is almost independent
of the cut-off angle $\theta_0$ for the range $20\,^{\circ} < \thmin < 40\,^{\circ}$.
We consider an optimal value of $30\,^{\circ}$, with the sensitivity on
Im $h^{Z,\gamma}_1$ being $1 \times 10^{-5}$, and Im $h^{Z,\gamma}_2$ being
$2.2 \times 10^{-4}$. 

\begin{figure}
\begin{minipage}{0.45\linewidth}
  \centering
  \includegraphics[width=6.5cm, height=5cm]{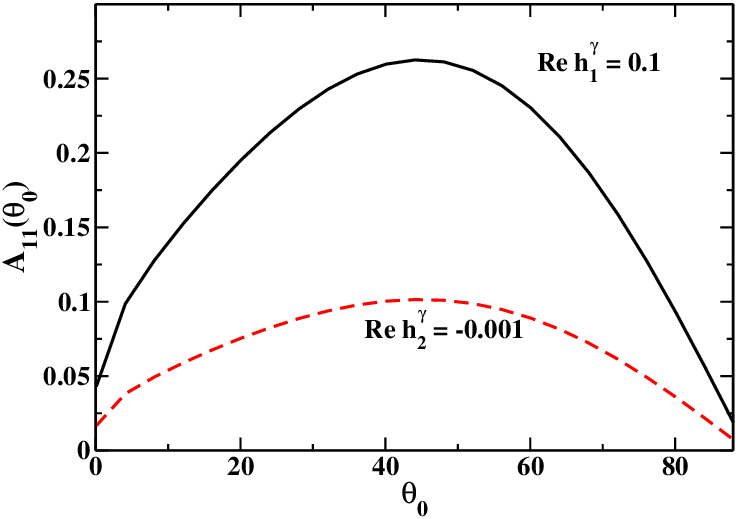}
  \caption{The asymmetry  $A_{11} (\thmin)$ defined in Eq.~(\ref{diff_asym_cpodd11}) as a function
of the cut-off angle $\theta_0$ for the various couplings.}
  \label{fig:asy11}
\end{minipage}%
\hspace{0.5cm}
\begin{minipage}{0.45\linewidth}
  \centering
  \includegraphics[width=6.5cm, height=5cm]{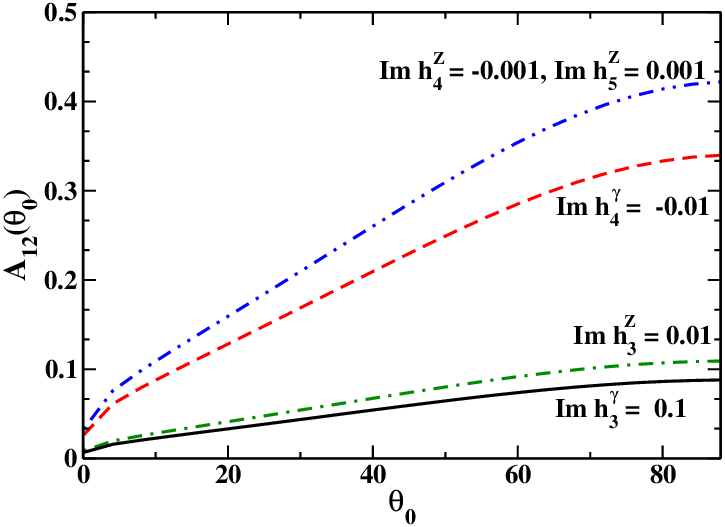}
  \caption{The asymmetry  $A_{12} (\thmin)$ defined in Eq.~(\ref{diff_asym_cpeven12}) as a function
of the cut-off angle $\theta_0$ for the various couplings.}
  \label{fig:asy12}
\end{minipage}
\end{figure}

\begin{figure}
\vspace{0.4cm}
\begin{minipage}{0.45\linewidth}
  \centering
  \includegraphics[width=6.5cm, height=5cm]{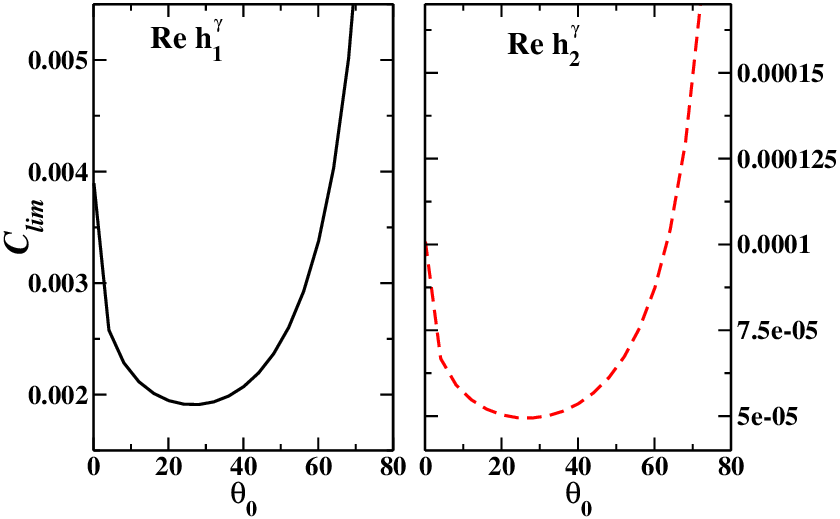}
  \caption{The 90\% CL limits on the anomalous couplings from  $A_{11} (\thmin)$, as a function
of the cut-off angle $\theta_0$.}
  \label{fig:asy11_lim}
\end{minipage}%
\hspace{0.5cm}
\vspace{0.4cm}
\begin{minipage}{0.45\linewidth}
  \centering
  \includegraphics[width=6.5cm, height=5cm]{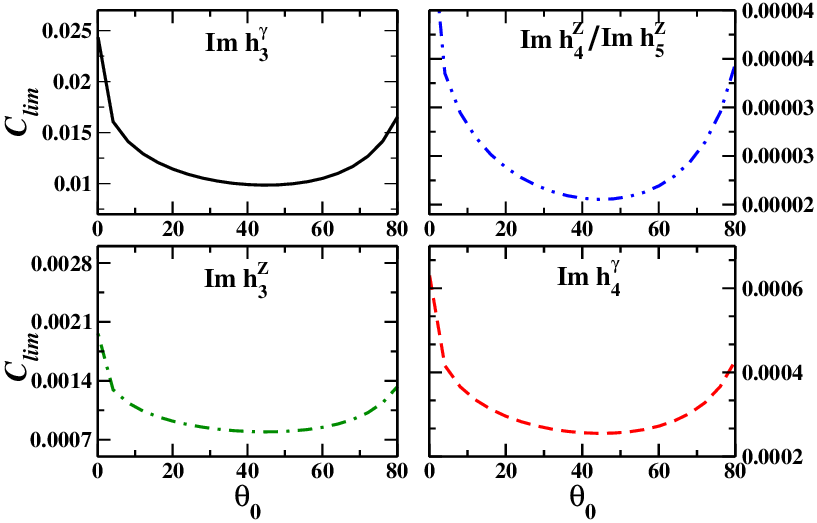}
  \caption{The 90\% CL limits on the anomalous couplings from  $A_{12} (\thmin)$, as a function
of the cut-off angle $\theta_0$.}
  \label{fig:asy12_lim}
\end{minipage}
\end{figure}

We next consider the asymmetries $A_{11,12} (\theta_0)$, which
are dependent on a different set of couplings. We would like to repeat that the SM
contribution to these asymmetries is zero. We plot in Figs.~\ref{fig:asy11}
and~\ref{fig:asy12}
the various asymmetries as a function of the cut-off angle $\thmin$. Each coupling
is set to a value such that the linear approximation holds good 
while the other couplings are set to zero. The 90\% CL limits obtained
on the various
couplings from these asymmetries are next shown in Figs.~\ref{fig:asy11_lim} and~\ref{fig:asy12_lim}.
Similar to the previous case, we see that the limits obtained are 
independent of $\thmin$
in the range $20\,^{\circ} < \thmin < 40\,^{\circ}$ in the case of $A_{11} (\thmin)$. We again
consider the optimal value of $30\,^{\circ}$, with Re $h^\gamma_1$ < 2 $\times 10^{-3}$
and Re $h^\gamma_2$ < 0.5 $\times 10^{-4}$. In case of $A_{12} (\thmin)$,
as can be seen from Fig.~\ref{fig:asy12_lim}, the limits on Im $h^{\gamma}_{3,4}$
and Im $h^Z_{3}$ have stable values over a wide range of $20\,^{\circ} < \thmin < 60\,^{\circ}$,
with the respective limits being Im $h^\gamma_3 < 1 \times 10^{-2}$, Im $h^\gamma_4 < 2.1 \times 10^{-4}$
and Im $h^Z_3 < 0.9 \times 10^{-3}$. The best limit
on Im $h^Z_{4,5}$ is $2.1 \times 10^{-5}$ for $\thmin = 40\,^{\circ}$.

\begin{figure}
\begin{minipage}{0.45\linewidth}
  \centering
  \includegraphics[width=6.5cm, height=5cm]{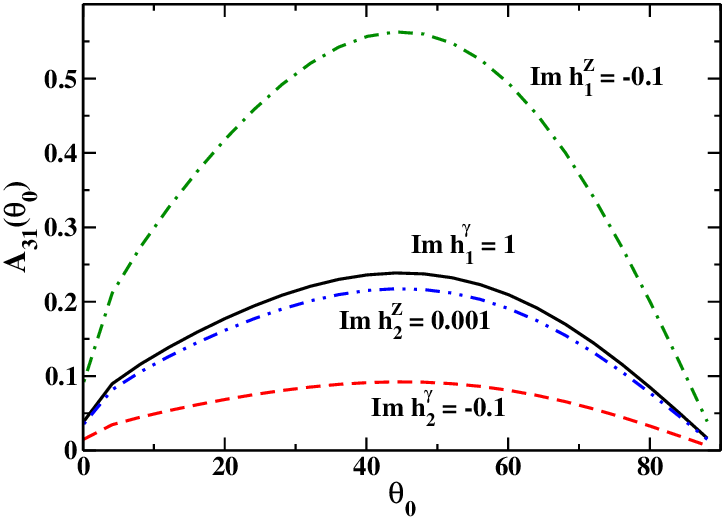}
  \caption{The asymmetry  $A_{31} (\thmin)$ defined in Eq.~(\ref{diff_asym_cpodd31}) as a function
of the cut-off angle $\theta_0$ for the various couplings.}
  \label{fig:asy41}
\end{minipage}%
\hspace{0.5cm}
\begin{minipage}{0.45\linewidth}
  \centering
  \includegraphics[width=6.5cm, height=5cm]{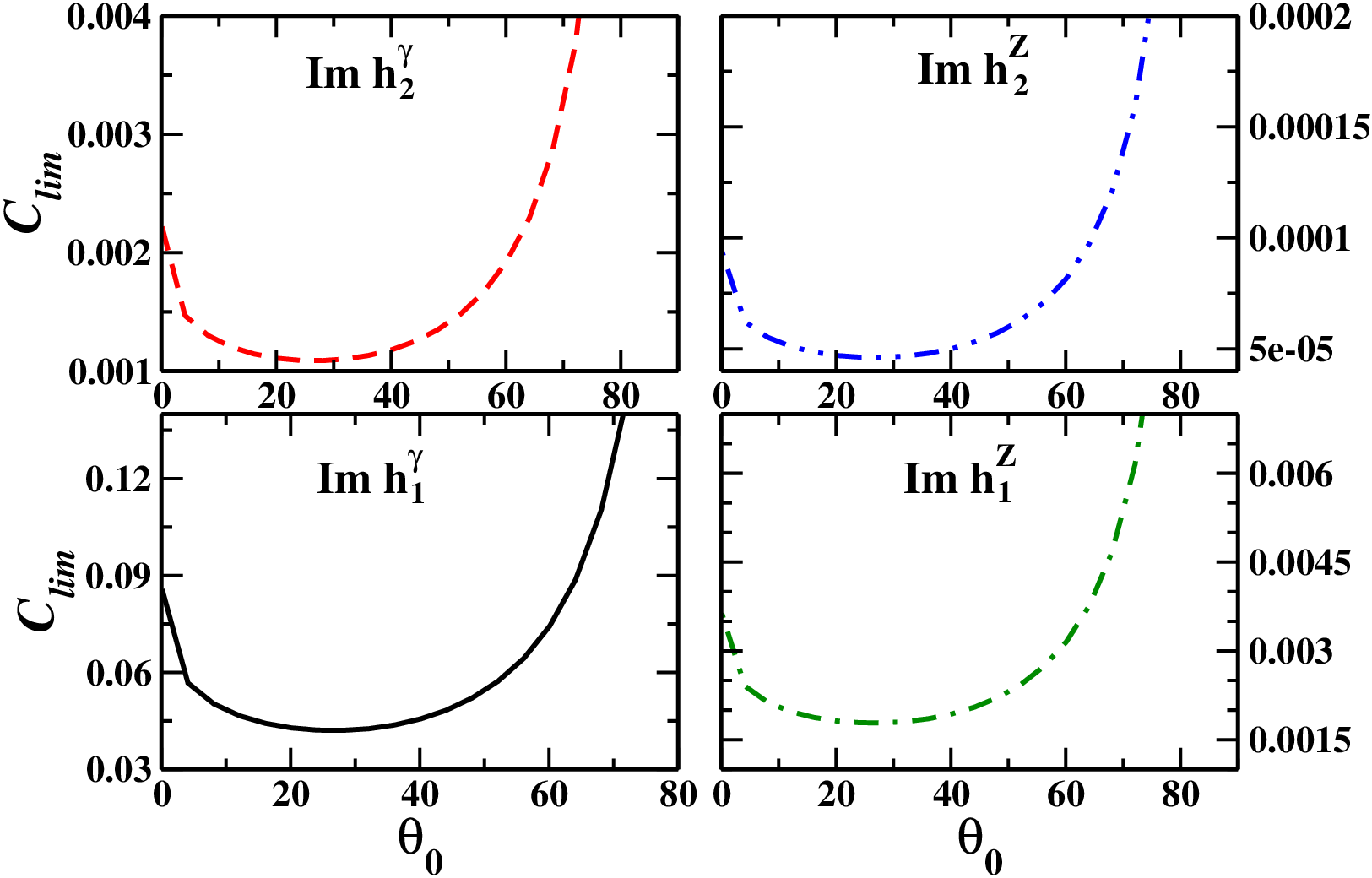}
  \caption{The 90\% CL limits on the anomalous couplings from  $A_{31} (\thmin)$, as a function
of the cut-off angle $\theta_0$.}
  \label{fig:asy41_lim}
\end{minipage}
\end{figure}

\begin{figure}
\vspace{0.4cm}
\begin{minipage}{0.45\linewidth}
  \centering
  \includegraphics[width=6.5cm, height=5cm]{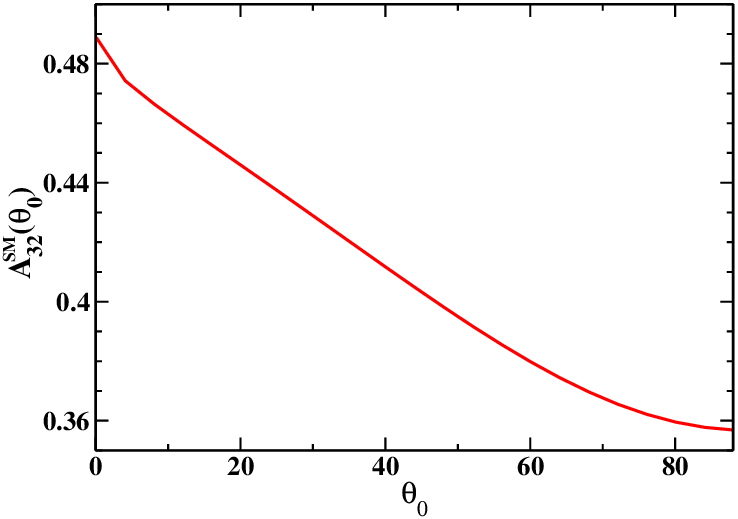}
  \caption{The SM dependent part $A^{SM}_{32} (\thmin)$ of the asymmetry $A_{32} (\thmin)$ 
  as a function of the cut-off angle $\theta_0$.}
  \label{fig:asy32_SM}
\end{minipage}%
\hspace{0.5cm}
\vspace{0.4cm}
\begin{minipage}{0.45\linewidth}
  \centering
  \includegraphics[width=6.5cm, height=5cm]{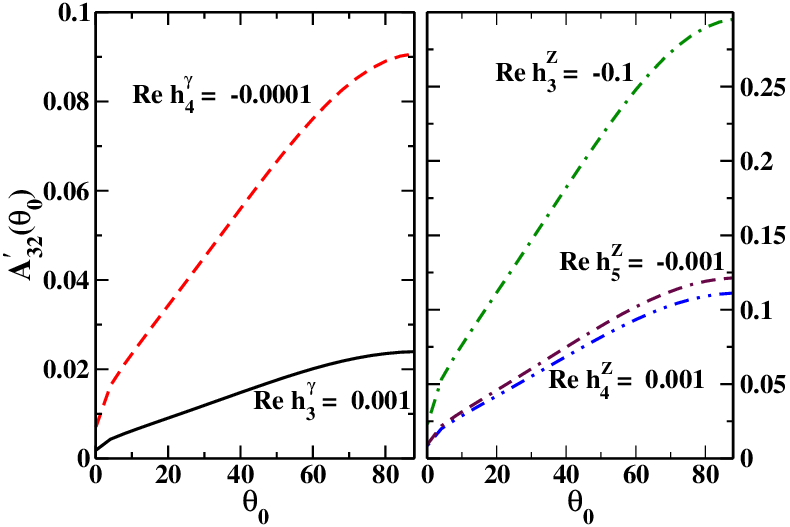}
  \caption{The asymmetry  $A'_{32} (\thmin)$  as a function
of the cut-off angle $\theta_0$ for the various couplings.}
  \label{fig:asy42}
\end{minipage}
\end{figure}

\begin{figure}
\centering
\includegraphics[width=7cm, height=5cm]{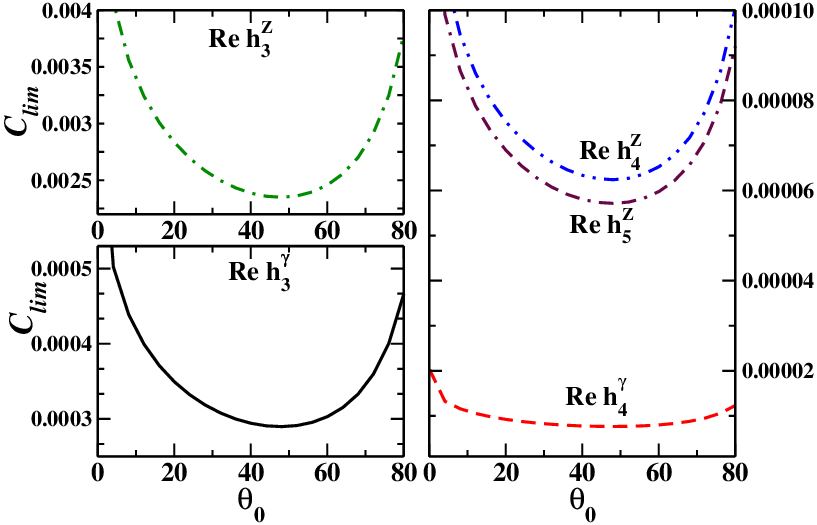}
\caption{The 90\% CL limits on the anomalous couplings from  $A'_{32} (\thmin)$, as a function
of the cut-off angle $\theta_0$.}
 \label{fig:asy42_lim}
\end{figure}

Finally, we present our results for the asymmetries $A_{31,32} (\thmin)$.
The asymmetry $A_{31}(\thmin)$ as a function of $\theta_0$, for the
various
couplings is shown in Fig.~\ref{fig:asy41}, with the 90\% CL limits on the couplings
from this asymmetry shown in Fig.~\ref{fig:asy41_lim}. The asymmetry $A_{32} (\thmin)$
contains the SM contribution $A^{SM}_{32} (\thmin)$ in addition to the contribution from anomalous couplings, so we only plot the  
contribution from the anomalous couplings defined as 
$A'_{32} (\thmin) = |A_{32} (\thmin) - A^{SM}_{32} (\thmin)|$. We then determine the 
individual 90\% CL limits on the couplings from $A_{32} (\thmin)$,
using the expression
\begin{equation}\label{limit_2}
\mathcal{C}_{lim} \equiv \frac{1.64 \sqrt{1-(A^{SM}_{32})^2}}{|A'_{32}| \sqrt{N_{SM}}}, 
\end{equation}
where $A'_{32}$ in the denominator is the value of $A'_{32} (\thmin)$
for unit value of the coupling.
The SM contribution to the asymmetry $A_{32} (\thmin)$ is shown in Fig.~\ref{fig:asy32_SM},
and the additional contribution to $A_{32} (\thmin)$, due to the different couplings apart from the 
SM, defined as $A'_{32} (\thmin)$ is shown in Fig.~\ref{fig:asy42}. 
The 90\% CL limits obtained on the couplings contributing to $A_{32} (\thmin)$ 
from Eq.~(\ref{limit_2}) is shown in Fig~\ref{fig:asy42_lim}.
We only present the result for this case, because the couplings which
enter $A_{21,22}(\thmin)$
are also present in $A_{31,32} (\thmin)$.
It can be seen from Eq.~(\ref{diff_asym_cpodd21}), that $A_{21}(\theta_0)$
receives contribution from the couplings Im $h^Z_{1,2}$ and Im $h^\gamma_{1,2}$,
whereas Re $h^\gamma_{3,4}$ contributes to $A_{22}(\theta_0)$, Eq.~(\ref{diff_asym_cpeven22}).
As these anomalous
couplings also contribute to the other asymmetries, and we have checked that
the individual limits obtained on these couplings from these asymmetries
are of the same order or better than the individual limits obtained from
$A_{21,22}(\theta_0)$. Therefore we do not show the results for these
asymmetries, but we list in Table~\ref{tab:asym2122_lim} the individual
limits obtained in this case.
\begin{table}
\begin{center}
\begin{tabular}{||c|c|c|c||c|c||}\hline\hline
\multicolumn{4}{||c||} {$A_{21}$}& 
\multicolumn{2}{||c||} {$A_{22}$} \\ \hline \hline
Im $h^\gamma_1$ & Im $h^\gamma_2$ & 
Im $h^Z_1$ & Im $h^Z_2$ &Re $h^\gamma_3$ & Re $h^\gamma_4$   \\ \hline
$2 \times 10^{-1}$ &$3 \times 10^{-4}$ &$1.5 \times 10^{-3}$ &$3 \times 10^{-5}$
&$8 \times 10^{-4}$ &$2 \times 10^{-5}$ \\ \hline\hline
 \end{tabular}
\caption{Table of sensitivities obtainable at the ILC with the
machine and operating parameters given in the text for
the asymmetries $A_{21}$ and $A_{22}$.}
\label{tab:asym2122_lim}
\end{center}
\end{table}
Finally we show in Table~\ref{tab:asym4142_lim} the best individual limits obtained  
from the asymmetry $A_{31,32} (\thmin)$.  The limits obtained get better with
increase in centre-of-mass energy.
\begin{table}
\begin{center}
\begin{tabular}{||c|c||c|c|c||}\hline\hline
\multicolumn{2}{||c||} {$A_{31}$} &
\multicolumn{3}{||c||} {$A_{32}$}  
\\ \hline \hline
Im $h^Z_1$ & Im $h^Z_2$  
&Re $h^Z_3$ &Re $h^Z_4$ & Re $h^Z_5$     \\ \hline 
$1 \times 10^{-3}$ & $2.5 \times 10^{-5}$  
&$2 \times 10^{-3}$ &$6 \times 10^{-5}$ & $5 \times 10^{-5}$   \\ \hline \cline{1-4} 
Im $h^\gamma_1$ & Im $h^\gamma_2$ & 
Re $h^\gamma_3$ & Re $h^\gamma_4$   \\  \cline{1-4} 
$2 \times 10^{-2}$ & $5 \times 10^{-4}$ &
$3 \times 10^{-4}$ & $8 \times 10^{-6}$  \\ \cline{1-4}  \cline{1-4} 
 \end{tabular}
\caption{Table of limits on couplings obtainable at the ILC with the
machine and operating parameters given in the text for
the asymmetries $A_{31}$ and $A_{32}$.}
\label{tab:asym4142_lim}
\end{center}
\end{table}

\section{Discussion and Conclusions}\label{sec:conclusion}

The gauge sector of the SM is one of the key corners which can provide a window into BSM physics.  
It has been one that has been studied extensively in the literature.  
It has also been probed to high precision at the LEP as well as at the LHC and 
Fermilab experiments.  Anomalous triple gauge boson couplings constitute 
an interesting and 
important  model-independent method by which BSM physics has been introduced.  
Another less popular but equally compelling method is to introduce BSM physics 
through contact interactions.  In fact, this latter has not received sufficient 
attention in the literature.

One of the missions of the present work is to explore whether anomalous 
TGCs capture all the essence of BSM physics, or whether one needs to go beyond that.  
Before embarking on this, we first asked ourselves if the anomalous TGCs 
considered in the literature are exhaustive or not.  It turns out, surprisingly, 
that from the considerations of Bose symmetry, gauge invariance, etc., it is 
possible to generate a term that has not been found in the literature.
One of the reasons could be that this term is not one that is invariant under 
the symmetry $Z \leftrightarrow \gamma$.
We find a $ZZ\gamma$ coupling, while there is no analogous $Z\gamma\gamma$ term.

While the bounds obtained in~\cite{Aad:2013izg, Chatrchyan:2013nda} might continue to be
valid approximately, the analysis of the data clearly would have
to be done afresh for more precise bounds in view of the above.

LHC experiments obtain bounds on TGCs by looking at the 
transverse momentum spectrum of the photon. 
Since the photon energy spectrum has similar sensitivities
to CP-violating and CP-conserving couplings, the LHC cannot 
discriminate between these couplings. Their results
are interpreted in terms of the CP-conserving couplings.

The CP-violating couplings
$h_{1,2}^{Z, \gamma}$ can be bounded by
studying CP-violating asymmetries, the simplest being the
forward-backward asymmetry of the photon in the type of
experiments performed at Tevatron. 
The corresponding effect in 
$e^+e^-\rightarrow \gamma Z$ was studied in~\cite{SDR}.
We have carried out a detailed numerical study of the implications of such BSM physics.
We have considered a list of asymmetries, in the presence of both
transverse and longitudinal polarization so as to give individual
limits on the CP-conserving and the CP-violating
 couplings. These asymmetries will help 
to discriminate between the CP-conserving and the CP-violating
 couplings. Moreover
we find that the limits obtained on the TGCs from the various
asymmetries will be better than those obtained from the LHC, and 
will improve with the centre-of-mass energy. In the presence 
of LP, we find the limits Im $h^{Z,\gamma}_1 < 1 \times 10^{-5}$
and Im $h^{Z,\gamma}_2 < 2.2 \times 10^{-4}$. The limits on the
other anomalous couplings are obtained in the presence of TP and are
listed in Sec.~\ref{sec:results} as well as in
Tables~\ref{tab:asym2122_lim} and \ref{tab:asym4142_lim}.

The two dimension-8 anomalous couplings pertaining to the $ZZ\gamma$ vertex,
$h^Z_{4,5}$ show similar behaviour in case of the various asymmetries. 
At a fixed energy, it turns out that the distributions are such that
the angular behaviour is the same. This is true in the case considered
in this work, which is one where the polarization of the two final-state
bosons is summed over. It is therefore important to discuss the matter
of discriminating between these two anomalous couplings.
If it is possible to have an energy scan at
the ILC, then the energy dependence would reveal whether the BSM
physics is due to $h^Z_4$ or due to $h^Z_5$.  

Alternatively, as 
in our previous work~\cite{Ananthanarayan:2011fr}, if the spin of the 
$Z$ is resolved, it is likely to lead to a situation
where one may be able to discriminate between the two sources,
since the vertices are actually different.  
It is clear from Table~\ref{tab:red_form} that the two-index tensor $\Gamma^Z_{\alpha \beta}$ 
is different for $h^Z_4$ and $h^Z_5$.
Summing over polarizations leads to the same $\theta$ and $\phi$ distributions.
Therefore, in order to probe the full tensor structure, it is necessary
to resolve the spin(s) of the boson(s).
The  $Z$ polarization vector,
when contracted with the $h_4^Z$ term in Eq.~(\ref{hag_1}) or the
$h_2^\gamma$ term in Eq.~(\ref{hag_2}), gives a factor $q.\epsilon (k_2)$. 
In the centre-of-mass frame, $q$ has only the time component, whereas 
for transverse polarization, $\epsilon (k_2)$ has no time component, the
corresponding amplitude vanishes. 
So only longitudinal polarization for $Z$ contributes to $h^Z_4$ (or
$h_4^\gamma$). 
On the other hand, both longitudinal and transverse $Z$ polarizations
would survive for the
$h^Z_5$ term. 
It is thus plausible that observing $Z$ polarization can distinguish between 
$h^Z_4$ and $h^Z_5$.
This is beyond the scope of the present work. 

In order to carry out a detailed comparison, we started out by reducing 
the familiar set of contact interactions to the anomalous TGCs.
While the TGCs in the case of CP-conserving
 interactions were expressed in terms of Levi-Civita 
terms, and the contact interactions without, we had to carry out a detailed exercise to 
carry out the comparison.
We have established a relation between these two approaches. While doing the analysis
we found that a triple gamma term ($r_7$) which has appeared only once in the literature
plays a definitive role. We also found that $r_7$ has no definite
CP transformation property, i.e. the operator multiplying $r_7$ is partly
CP odd and partly CP even.
Our conclusions are that anomalous  TGC terms do not exhaust all possible 
distributions that can be generated by contact interactions.

Although our work is motivated by the immediate goal of finding a 
detailed physics programme for the ILC, it has a more general import.
These may be listed as follows: \\
($a$) A general analysis of the physics of gauge bosons in a model-independent manner, 
subject only to the constraints of gauge invariance and Lorentz invariance.
This is obviously of importance also to the LHC. \\
($b$) It is of importance to the Compact Linear Collider (CLIC)~\cite{Abramowicz:2013tzc}  which also requires
a dedicated physics programme, lot of which would be common to the ILC.
In the coming years, many of these analyses could be done for CLIC energies
and polarization. There would be many distinguishing features between the two 
as regards the detector capabilities, which are beyond the scope for the present paper.

\bigskip

\noindent {\bf{Acknowledgements:}} 
SDR thanks the Department of Science and Technology, Government of
India, for support under the J.C. Bose National Fellowship program,
grant no. SR/SB/JCB-42/2009.  SDR also thanks Prof. Debajyoti Choudhury
for interesting discussions and collaboration at early stages of
this project. MP thanks Physical Research Laboratory, Ahmedabad for its hospitality where
part of this work was done.

\appendix

\section{Conversion of anomalous TGCs involving the Levi-Civita symbol}\label{appendix:reduction}

As can be seen from Table~\ref{tab:red_form},
some of the  anomalous TGC couplings 
involve the Levi-Civita symbol. The contact interactions
discussed in Ref.~\cite{BASDR1, BASDR2} however do not involve these symbols.
Therefore it will be useful to convert the Levi-Civita symbols to
a form equivalent to that used for contact interactions only involving
the momentum four vectors and the Dirac matrices. We therefore present 
below a derivation of 
simplified forms for the anomalous couplings involving $h^V_{3,4,5}$
containing Levi-Civita symbols. 

Firstly we would like to observe that the $\slashed{q}$ terms, occurring
singly, 
can be dropped, because they give $0$ on 
using the Dirac equation for the electron and positron
spinors: 
\begin{equation}
\bar v(p_+)\slashed{q}u(p_-)=\bar
v(p_+)(\slashed{p}_-+\slashed{p}_+u(p_-) = 0.
\end{equation}
\noindent 
The terms with $\slashed{q}\gamma_5$ give zero in the limit of
vanishing electron mass, and can also be dropped.

We now take up various terms in Table 1 containing the Levi-Civita
tensor by turns. At all stages, we set $k_1^{\alpha}$, $k_2^{\beta}$ and
$q^\nu\equiv (k_1 + k_2)^\nu\equiv (p_+ + p_-)^\nu$ to $0$.

\noindent 1. The $h_3^{\gamma,Z}$ terms have $\Gamma^{\alpha \beta}=\gamma_{\nu}
\epsilon^{\alpha \beta \nu k_1}$.
In this term, we can introduce the identity
\beq\label{g5identity}
1= \gamma_5^{2} =-\frac{i}{4!} \gamma_5 
\epsilon^{\rho \lambda \sigma \tau}\gamma_{\rho} 
\gamma_{\lambda} \gamma_{\sigma} \gamma_{\tau}
\eeq
to get 
\beq
\Gamma^{\alpha \beta} = \frac{i}{4!} \gamma_5 \gamma_{\nu} \gamma_{\rho} \gamma_{\lambda}
\gamma_{\sigma} \gamma_{\tau} \epsilon^{\alpha \beta \nu k_1}
\epsilon^{\rho \lambda \sigma \tau}.
\eeq
Of the 5 indices $\nu, \rho, \sigma, \lambda, \tau$ in $d=4$ at least 2 indices have 
to be equal. Because of the presence of the totally antisymmetric
$\epsilon^{\rho \sigma \lambda \tau}$, only $\nu$ will be allowed
to be equal to one of $\rho, \sigma, \lambda, \tau$ giving 4 equal terms. Then 
\begin{equation}
\Gamma^{\alpha \beta}= \frac{i}{3!} \gamma_5 \gamma_{\lambda} \gamma_{\sigma} 
\gamma_{\tau}\epsilon_{\nu}^{\lambda \sigma \tau} \epsilon^{\alpha \beta \nu k_1}
\end{equation}
We now use the identity
\beq
\begin{array}{rcl}
\epsilon_{\nu}^{\lambda \sigma  \tau} \epsilon^{\alpha \beta \nu k_1 }
&=& g^{\lambda \alpha}(g^{\sigma \beta} k_1^{\tau} - g^{\tau \beta}
k_1^{\sigma})
- g^{\lambda \beta} (g^{\sigma \alpha} k_1^{\tau} - g^{\tau \alpha}
  k_1^{\sigma}) \\
&&- k_1^{\lambda} (g^{\sigma \beta} g^{\alpha \tau} - g^{\tau \beta} g^{\alpha \sigma}) 
\end{array}
\eeq
to get
\begin{equation}
\Gamma^{\alpha \beta} 
=-i \gamma_5(\slashed{k}_1\gamma^{\alpha} \gamma^{\beta}  + \gamma^{\alpha} k_1^{\beta} 
- \slashed{k}_1 g^{\alpha \beta}) \label{h3_1}
\end{equation}

\noindent 2. The $h_4^{\gamma,Z}$ terms have
\beq
\Gamma^{\alpha \beta} = -
\gamma_{\nu} k_1^{\beta} \epsilon^{\alpha \nu k_1 q}
\eeq

As before using the identity in Eq. \ref{g5identity}, we can write
\beq
\Gamma^{\alpha \beta}=
-\frac{i}{4!}\gamma_5 \gamma_{\nu} \gamma_{\rho} \gamma_{\sigma} \gamma_{\lambda} 
\gamma_{\tau} \epsilon^{\rho \sigma \lambda \tau} k_1^{\beta}
\epsilon^{\alpha \nu k_1 q},
\eeq
which on  using the fact that one pair of 5 indices has to be equal,
gives
\begin{eqnarray}\label{h4_1}
\Gamma^{\alpha \beta} &=& -\frac{i}{3!} \gamma_5 \epsilon_{\nu}^{\;\sigma \lambda \tau} \gamma_{\sigma}
\gamma_{\lambda} \gamma_{\tau} k_1^{\beta} \epsilon^{\alpha \nu k_1 q} 
\end{eqnarray}
This equation reduces, on using the identity for a product of two
Levi-Civita tensors to be written in terms of Kronecker deltas, to 
\begin{eqnarray}\label{h4_2}
\Gamma^{\alpha \beta}
 &=& -\frac{i}{3!2} \gamma_5 k_1^{\beta} (\gamma^{\alpha} \slashed{k}_1 \slashed{q}
- \gamma^{\alpha} \slashed{q} \slashed{k}_1 - \slashed{k}_1 \gamma^{\alpha} \slashed{q} 
+ \slashed{k}_1 \slashed{q} \gamma^{\alpha} + \slashed{q} \gamma^{\alpha} \slashed{k}_1
- \slashed{q} \slashed{k}_1 \gamma^{\alpha})
\end{eqnarray}
We now use $\slashed{q}=\slashed{p}_-+\slashed{p}_+$, then anti-commute $\slashed{p}_-$ to the extreme
right and $\slashed{p}_+$ to the extreme left,
 and use $\slashed{p}_- u(p_-)=0;~\bar{v}(p_+)\slashed{p}_+=0$.
Dropping $k_1^{\alpha}$ terms because $(\epsilon_{\alpha}(k_1)
k_1^{\alpha}=0)$,
the result is 
\begin{equation}
\Gamma^{\alpha \beta}= -i \gamma_5 k_1^{\beta} (p_+ \cdot k_1 \gamma^{\alpha} - p_- \cdot k_1 \gamma^{\alpha} -
p_+^{\alpha}\slashed{k}_1 + p_-^{\alpha}\slashed{k}_1)
\end{equation}

\noindent 3. The $h_5^{Z}$ term is 
\begin{equation}\label{h6_1}
\Gamma^{\alpha \beta}= 
\gamma^{\nu}2 k_2^{\alpha} \epsilon^{\beta k_2 \nu q} - \frac{(s -
m_Z^2)}{2} \gamma^{\nu}(\epsilon^{\alpha \beta k_2 \nu} 
+\epsilon^{\alpha \beta q \nu})
\end{equation} 
The first term of Eq.~\ref{h6_1}, following the same procedure as before
yields
\beq
\label{h6_2}
\begin{array}{rcl}
2 \gamma_{\nu} k_2^{\alpha} \epsilon^{\beta k_2 \nu q}
\!&=&\!\! - \displaystyle  \frac{4i}{3!} k_2^{\alpha} \gamma_5 
\left[\gamma_{\beta} (\slashed{k}_2 \slashed{q} 
- \slashed{q} \slashed{k}_2) - \slashed{k}_2 \gamma_{\beta} \slashed{q} + \slashed{q} 
\gamma_{\beta} \slashed{k}_2 - 
\slashed{q} \slashed{k}_2 \gamma_{\beta} + \slashed{k}_2 \slashed{q} \gamma_{\beta}\right].
\end{array}
\eeq
Now  using $\slashed{q}=\slashed{p}_-+\slashed{p}_+$ and then commuting
$\slashed{p}_+$ through to the extreme left and $\slashed{p}_-$ to the
extreme right and using the Dirac equation, the right-hand side of 
Eq.~\ref{h6_2} becomes
\begin{eqnarray}\label{h6_3}
&& 2 i \gamma_5[(p_- \cdot k_2 - p_+ \cdot k_2) \gamma_{\beta} - (p_--p_+)_{\beta}\slashed{k}_2] k_2^{\alpha}
\end{eqnarray}
The second term of Eq.~\ref{h6_1} simplifies to 
\begin{eqnarray}\label{h6_4}
\gamma^{\nu} \epsilon^{\alpha \beta k_2}_{\nu} &=& 
 i \gamma_5 \left[\gamma^{\alpha} \gamma^{\beta} \slashed{k}_2 
- g^{\alpha \beta} \slashed{k}_2 + k_2^{\alpha} \gamma^{\beta}\right]  
\end{eqnarray}
Similarly the third term of Eq.~\ref{h6_1} is
\begin{eqnarray}\label{h6_5}
\gamma_{\nu} \epsilon^{\alpha \beta q \nu} 
&=&  i  \gamma_5 \left[(p_+-p_-)^{\beta} \gamma^{\alpha} + (p_- -p_+)^{\alpha} \gamma^{\beta}\right] 
\end{eqnarray}
Combining Eqs.~\ref{h6_3},~\ref{h6_4} and~\ref{h6_5}, $V_{\alpha
\beta}$ for the $h_5^{Z}$ term  
takes the form
\begin{eqnarray}\label{h6_6}
&&i\gamma_5 \left[ 2 k_2^{\alpha} (p_--p_+)\cdot
k_2 \gamma^{\beta} -2 k_2^{\alpha}(p_--p_+)^{\beta}\slashed{k}_2
\right. \nonumber \\
&& \left. 
+ \frac{(s-m_Z^2)}{2} \left\{(\gamma^{\alpha} \gamma^{\beta} \slashed{k}_2 
- g^{\alpha \beta}\slashed{k}_2 + k_2^{\alpha} \gamma^{\beta}) \right. \right. \nonumber \\
&& \left. \left.
+ ((p_--p_+)^{\alpha} \gamma^{\beta}-(p_--p_+)^{\beta} \gamma^{\alpha})\right\}\right]
\end{eqnarray}
In Eq.~\ref{h6_6}, $\slashed{k}_2$ can be replaced by $\slashed{k}_1 $ 
using
$\slashed{k}_2 = \slashed{q} - \slashed{k}_1$.
Then, using, as before, $\slashed{q}=\slashed{p}_-+\slashed{p}_+$, and
the Dirac equation after appropriately commuting through $\slashed{p}_-$
and $\slashed{p}_+$ to the extreme right and left, respectively, we get
\beq
\begin{array}{rcl}
\Gamma^{\alpha \beta}
&=&  i \gamma_5 \left[ 
2 k_2^{\alpha} \left\{ - k_1 \cdot (p_--p_+) \gamma^{\beta} 
+ (p_--p_+)^{\beta} \slashed{k}_1
\right\}
 \right.  \\
&+& \left. 
 \frac{(s-m_Z^2)}{2}  \left(- \slashed{k}_1\gamma^{\alpha}  \gamma^{\beta} + k_1^{\beta} \gamma^{\alpha} - 
+ 2 (p_--p_+)^{\alpha} \gamma^{\beta}
g^{\alpha \beta} \slashed{k}_1 -
2 (p_--p_+)^{\beta} \gamma^{\alpha}\right)\right]
\end{array}
\eeq



\end{document}